\documentstyle[12pt]{article}
\begin{document}
\vskip 2cm
\begin{center}
{\sf {\Large Nilpotent Symmetries of a 4$D$ Model of the Hodge Theory: Augmented (Anti-)Chiral Superfield Formalism}}

\vskip 2.0cm

{\sf  A. Shukla$^{(a)}$\footnote {{\it Current address:
 School of Physics and Astronomy, Sun Yat-Sen University, Zhuhai - 519082, China}}, N. Srinivas$^{(a)}$, R. P. Malik$^{(a,b)}$}\\
$^{(a)}$ {\it Physics Department (CAS), Institute of Science,}\\
{\it Banaras Hindu University (BHU), Varanasi - 221 005, (U.P.), India}\\

\vskip 0.5cm

$^{(b)}$ {\it DST Centre for Interdisciplinary Mathematical Sciences,}\\
{\it Institute of Science, Banaras Hindu University, Varanasi - 221 005, India}\\
{\small {\sf {e-mails: ashukla038@gmail.com; seenunamani@gmail.com;   rpmalik1995@gmail.com}}}

\end{center}

\vskip 1cm

\noindent
{\bf Abstract:} 
We derive the continuous nilpotent symmetries of the four (3 + 1)-dimensional (4$D$) model of the Hodge theory (i.e. 4$D$ Abelian 2-form gauge
theory) by exploiting the beauty and strength of the symmetry invariant restrictions on the (anti-)chiral superfields.
The above off-shell nilpotent symmetries are the Becchi-Rouet-Stora-Tyutin (BRST),
anti-BRST and (anti-)co-BRST transformations which turn up beautifully due to the (anti-)BRST and (anti-)co-BRST invariant restrictions 
on the (anti-)chiral superfields that are defined on the (4, 1)-dimensional (anti-)chiral super-submanifolds of the {\it general} (4, 2)-dimensional supermanifold on which our {\it ordinary} 4$D$ theory is generalized. The latter supermanifold is characterized by the superspace coordinates $Z^M = (x^\mu,\, \theta,\, \bar\theta)$ where $x^\mu\, (\mu = 0, 1, 2, 3 )$ are the bosonic coordinates and a pair of Grassmannian variables 
$\theta$ and $\bar\theta$ are fermionic in nature as they obey the standard relationships: $\theta^2 = {\bar\theta}^2 = 0,\, \theta\,\bar\theta + \bar\theta\,\theta = 0$). The derivation of the {\it proper} (anti-)co-BRST symmetries and proof of the absolute anticommutativity property 
of the conserved (anti-)BRST and (anti-) co-BRST charges are {\it novel} results of our present investigation 
(where {\it only} the (anti-)chiral superfields and their super-expansions have been taken into account).\\



\newpage

\noindent
\section {Introduction}
Superfield approach [1-8] to Becchi-Rouet-Stora-Tyutin (BRST) formalism is one of the very intuitive approaches which provides
the geometrical origin and interpretation for the nilpotency and absolute anticommutativity
properties of the (anti-)BRST symmetry transformations. The latter symmetries are the generalizations of a given local ``classical" gauge symmetry of a gauge theory to its counterparts ``quantum" symmetries which are known as BRST and anti-BRST symmetries. The very existence of the nilpotent (anti-)BRST symmetry transformations leads to the covariant canonical quantization of a given gauge theory because the canonical conjugate momenta exist for {\it all} the dynamical fields of an (anti-)BRST invariant gauge theory. Bonora-Tonin (BT) superfield formalism
[4, 5] has been quite successful in providing the derivation of {\it proper} (anti-)BRST  
transformations for the $p$-form ($p =1, 2, 3,...$) gauge theories
where the celebrated horizontality condition (HC) plays an important role. However, this condition (i.e. HC) leads to 
the derivation of proper (anti-)BRST symmetries {\it only} for the gauge and corresponding (anti-)ghost fields for a given 
$D$-dimensional {\it ordinary} gauge theory. The HC does {\it not} shed any light on the (anti-)BRST transformations
that are associated with the {\it matter} fields of a given {\it interacting} $p$-form gauge theory where there is a
precisely defined coupling between the $p$-form gauge field and matter fields.

In a set of papers [9-13], we have been able to generalize the BT-superfield formalism where, in addition to the HC, a set of gauge
invariant restrictions (GIRs) have  {\it also} been imposed on the superfields  to obtain the proper (anti-)BRST symmetry transformations 
for the gauge, (anti-)ghost and {\it matter} fields of an {\it interacting} gauge theory {\it together} (without spoiling the 
geometrical interpretation of the nilpotency and absolute anticommutativity properties in terms of the translational generators 
along the Grassmannian directions of the ($D$, 2)-dimensional supermanifold
on which the ordinary $D$-dimensional locally gauge invariant theory is generalized). In our earlier works [14-16], we have demonstrated
that any arbitrary Abelian $p$-form ($p = 1, 2, 3$) gauge theory (with Lorentz gauge-fixing term in the Feynman gauge) would also respect 
the (dual-)gauge symmetry transformations in $D = 2p$ dimensions of spacetime. As a consequence, one can obtain, in addition to the 
(anti-)BRST transformations, a proper (i.e. off-shell nilpotent and absolutely anticommuting) set of (anti-)dual-BRST [or (anti-)co-BRST]  
transformations for the above kind of Abelian $p$-form gauge theories (in $D = 2p$ dimensions of spacetime).
These theories turn out to be models for the Hodge theory where the symmetries and conserved charges provide the physical realizations of cohomological operators of differential geometry.

The purpose of our present investigation is to derive the {\it proper} (anti-)BRST transformations 
for the 4$D$ Abelian 2-form gauge theory {\it without} using the HC (which is primarily mathematical in nature). The results 
of our present endeavor, however, lend support emphatically to the preciseness of the results that have been obtained by using mathematically 
elegant and powerful  HC (see, e.g. [4, 5, 17]). Furthermore, we derive the proper (i.e. off-shell nilpotent and 
absolutely anticommuting) set of (anti-)co-BRST symmetry transformations which is a completely {\it novel} result in our 
present investigation. We exploit, in our present endeavor,  the (anti-)chiral superfields (defined
on the (4, 1)-dimensional super-submanifolds of the general (4, 2)-dimensional supermanifold) and invoke the (anti-)BRST and (anti-)co-BRST
invariant restrictions (BRSTIRs and CBRSTIRs) on the (anti-)chiral superfields to derive {\it all} the proper (anti-)BRST as well as (anti-)co-BRST
symmetry transformations for our present 4$D$ Abelian 2-form gauge theory. We further emphasize
that we have {\it not} used the restrictions on the (anti-)chiral superfields that
have their origin in the {\it mathematical} HC which has already been used to derive only
the (anti-)BRST symmetries within the framework of {\it usual} superfield formalism in our earlier work [17].

The central result of our present endeavor is the simplicity and generality of our technique which is useful in the derivation of nilpotent
(anti-) BRST as well as (anti-)co-BRST symmetry transformations. We provide the geometrical basis for the nilpotency property
which is equivalent to two successive translations of the superfields (derived after the application of BRSTIRs as well as
CBRSTIRs) along the Grassmannian directions of the super-submanifolds of the general (4, 2)-dimensional
supermanifold (on which our 4$D$ ordinary (anti-)BRST and (anti-)co-BRST invariant Abelian 2-form gauge theory is generalized).
We also furnish the geometrical interpretation for the (anti-)BRST as well as (anti-)co-BRST invariance of the appropriate
Lagrangian densities of our present theory. The proof of the absolute anticommutativity property of the conserved 
fermionic charges is a completely {\it novel} observation within the framework of augmented version of (anti-)chiral 
superfield approach to BRST formalism where {\it only} the (anti-)chiral expansions for the superfields have
been taken into account. In fact, this (i.e. the proof of the absolute anticommutativity of the (anti-)BRST charges) is the {\it highlight} of 
our present work on the (anti-)chiral superfield approach to BRST formalism.

The motivating factors behind our present investigation are as follows. First, we have derived the (anti-)BRST  
transformations by using the theoretical strength of symmetry invariance(s) ({\it without} any use of the HC). This is a 
novel result because we have used {\it only} the physical restrictions on the (anti-)chiral (super)fields which are inspired 
by the (anti-)BRST invariance. In other words, our present theoretical approach supports the precise results that have been obtained by 
using the mathematical strength of the HC [4, 5, 17]. Second, the derivation of the proper (anti-)co-BRST symmetries is {\it also} 
a novel result because, in our earlier work [17], we have {\it not} been able to achieve this goal within the framework of {\it usual}
superfield formalism. Third, we have developed a general technique which is useful in the derivation of {\it all} nilpotent 
symmetry transformations of our theory.  Fourth, the proof of the absolute anticommutativity properties of the (anti-)BRST and (anti-) co-BRST
{\it charges} is surprisingly a new result in view of the fact that we have taken into account {\it only} the (anti-)chiral super-expansions.
Finally, our present investigation is our first step towards our main 
goal to demonstrate that one can derive the {\it proper} (anti-) BRST and (anti-)co-BRST symmetries {\it together} 
based on the BRSTIRs and CBRSTIRs on the (anti-)chiral superfields that {\it do not} spoil the geometrical interpretations 
(of the very same symmetries which are derived from the application of mathematically powerful HC). 
Thus, our present investigation, although simple, strengthens and lends support to the results that are obtained by using the HC [17].

One of the key observations of our present endeavor is the fact the results of Sec. 5 (see below) are intertwined and inter-dependent in the sense that 
we have been able to express the fermionic charges (i.e. $Q_{(a)b},\, Q_{(a)d}$) in {\it exact} forms (using the CF-type of restrictions (cf. (7)),
and {\it vi\`ce-versa}. We have algebraically played with the expressions in the {\it ordinary} space and {\it superspace} which have helped
each-other in the derivation of equations (38), (40), (41), (43), (45)-(47). In other words, we have been able to prove the nilpotency
as well as absolute anticommutativity of the fermionic (anti-)BRST and (anti-)co-BRST charges due to our knowledge of superfield formalism
and properties of the fermionic symmetry transformations in the {\it ordinary} space. We draw the conclusion, ultimately, 
that the equations and contents of Sec. 5 are inter-dependent and inter-connected in an elegant manner.

The contents of our present paper are organized as follows. In Sec. 2, we very concisely mention the off-shell nilpotent 
(anti-)co-BRST and (anti-) BRST symmetry transformations within the framework of Lagrangian formalism. Our Sec. 3 deals 
with the derivation of (anti-)co-BRST transformations by using the (anti-)chiral superfields and imposing on them the symmetry motivated
(anti-)co-BRST invariant restrictions. The subject matter of Sec. 4 is the (anti-)BRST restrictions on the 
(anti-)chiral superfields that lead to the derivation of (anti-)BRST symmetry transformations. Our Sec. 5 contains the theoretical material on 
the proof of nilpotency and absolute anticommutativity of the 
fermionic (anti-) BRST and (anti-)co-BRST charges. Finally, we make some concluding remarks and point out a few future
theoretical directions for further investigations in our Sec. 6.

Our Appendix A deals with the derivation of the Curci-Ferrari (CF)-type restrictions by the requirement of absolute 
anticommutativity between the co-BRST and anti-co-BRST transformations as well as that of the
BRST and anti-BRST. The subject matter of Appendix B concerns itself with the {\it natural} proof of the property of
absolute anticommutativity and its connection with the {\it full} expansion of the superfields along the Grassmannian 
($\theta, \bar\theta$) directions of the $(4, 2)$-dimensional supermanifold.
 \vskip 0.5cm 
\noindent
{\it Convention and notations:} We adopt here the convention 
and notations such that the flat Minkowskian 4$D$ background spacetime manifold is endowed with a metric $\eta_{\mu\nu}$ 
which has signatures ($+1, -1, -1, -1$) so that the dot product between two non-null 4-vectors 
$P_\mu$ and $Q_\mu$ is: $P \cdot Q = \eta_{\mu\nu}\, P^\mu\, Q^\nu = P_0\, Q_0 - P_i\, Q_i$
where the Greek indices $\mu,\, \nu,\, \eta,\, \kappa,... = 0, 1, 2, 3$ and Latin indices $i,\,j,\,k,... = 1, 2, 3.$ 
The 4$D$ Levi-Civita tensor $\varepsilon_{\mu\nu\eta\kappa}$ satisfies $\varepsilon_{\mu\nu\eta\kappa}\,\varepsilon^{\mu\nu\eta\kappa} = -4!,\;
\varepsilon_{\mu\nu\eta\kappa}\,\varepsilon^{\mu\nu\eta\xi} = -3!\, \delta^\xi_\kappa,\,
\varepsilon_{\mu\nu\eta\kappa}\,\varepsilon^{\mu\nu\alpha\beta} = -2!\, (\delta^\alpha_\eta \,\delta^\beta_\kappa 
- \delta^\alpha_\kappa \,\delta^\beta_\eta),$ etc., and we choose $\varepsilon_{0123} = + 1 = - \varepsilon^{0123}.$ Throughout the whole
body of our text, we follow the notations $s_{(a)b}$ and $s_{(a)d}$ for the (anti-)BRST and (anti-)co-BRST symmetry transformations, respectively. 
The corresponding conserved and nilpotent charges are denoted by $Q_{(a)b}$ and $Q_{(a)d}$.

\vskip 0.5cm
\noindent
{\it Definition:} On a spacetime manifold without a boundary, we define a set of {\it three} cohomological
operators $(d, \delta, \Delta) $ where $d = dx^\mu\,\partial_\mu$ is the exterior derivative, 
$\delta = \pm \,*\,d\,*$ is the co-exterior derivative and $\Delta = (d + \delta)^2 = \{d, \delta\}$ is
the Laplacian operator. Here $*$ is the Hodge-duality operation on the above spacetime manifold. Together, these
operators satisfy the algebra: $d^2 = 0, \,\delta^2 = 0, \, \Delta = (d + \delta)^2 = d\delta + \delta d,\; [\Delta , d] = 0$
and $[\Delta , d] = 0$. Hence, the Laplacian operator $\Delta$ behaves like the Casimir operator (but not in the Lie algebraic sense).
To be precise, these operators are called as the de Rham cohomological operators of differential geometry (see, e.g. [18-21] for details).

\noindent
\section {Preliminaries: Lagrangian Formalism}
We very briefly mention here the off-shell nilpotent and absolutely anticommuting (anti-) BRST ($ s_{(a)b}$) 
and (anti-)co-BRST ($s_{(a)d}$) {\it symmetry} transformations for the following coupled Lagrangian densities 
$({\cal L}_{B,{\cal B}})$ and $({\cal L}_{{\bar B},\bar{\cal B}} )$ (see, e.g. [22, 23] for details)
\begin{eqnarray}
{\cal L}_{(B,\,\cal B)} & = & \frac{1}{2} {\cal B}\cdot{\cal B} - {\cal B}^{\mu} \Bigl(\frac{1}{2}\varepsilon_{\mu\nu\eta\kappa}\partial^{\nu}B^{\eta\kappa} 
+ \partial_{\mu}\phi_{2}\Bigr) +  B^{\mu}\Bigl(\partial^{\nu}B_{\nu\mu} + \partial_{\mu}\phi_{1}\Bigr) \nonumber \\ 
&-& \frac{1}{2}B \cdot B  + \partial_{\mu}\bar{\beta}\partial^{\mu}\beta + (\partial_{\mu}\bar{C_{\nu}} - \partial_{\nu}\bar{C_{\mu}})(\partial^{\mu}C^{\nu}) 
+  (\partial \cdot C - \lambda)\,\rho \nonumber\\
&+& (\partial \cdot \bar{C} + \rho)\,\lambda,
\end{eqnarray}
\begin{eqnarray}
{\cal L}_{(\bar{B},\, \bar{\cal{B})}} & = & \frac{1}{2}\bar{\cal B}\cdot \bar{\cal{B}}
 - \bar{\cal B}^{\mu}\Bigl(\frac{1}{2}\varepsilon_{\mu\nu\eta\kappa}\partial^{\nu}B^{\eta\kappa} - \partial_{\mu}\phi_{2} \Bigr)
 +  {\bar B}^{\mu}\Bigl(\partial^{\nu}B_{\nu\mu} - \partial_{\mu}\phi_{1} \Bigr) \nonumber\\
&-& \frac{1}{2}\bar B\cdot \bar{B}  + \partial_{\mu}\bar{\beta}\partial^{\mu}\beta  +  (\partial_{\mu}\bar{C_{\nu}} 
- \partial_{\nu}\bar{C_{\mu}})(\partial^{\mu}C^{\nu})  
 +  (\partial \cdot C - \lambda)\,\rho \nonumber\\
&+& (\partial \cdot \bar{C} + \rho )\,\lambda,
\end{eqnarray}
which describe the {\it free} 4$D$ Abelian 2-form gauge theory\footnote {It should be noted that we slightly
differ from the Lagrangian densities that have been taken in [23]. We do {\it not} have here a factor of $\frac{1}{2}$ associated with
fields $\phi_1$ and $\phi_2$ in our earlier work [23]. This is why, the (anti-)BRST and (anti-)co-BRST symmetry
transformations for these fields are slightly {\it different} in Eqns. (3) to (6).}. In the above, the 4$D$ vector fields 
(${\cal B}_\mu,\, \bar{\cal B}_\mu, \, B_\mu,\, {\bar B}_\mu$) are the Nakanishi-Lautrup type auxiliary fields, 
$({\bar C}_\mu) C_{\mu}$ are the {\it fermionic} ($ C^2_{\mu} = {\bar C}^2_\mu = 0, C_\mu\,{\bar C}_\nu + {\bar C}_\nu\, C_\mu = 0,$ etc.) 
Lorentz vector (anti-)ghost fields, $\bar{(\beta)}\beta $ are the {\it bosonic} Lorentz scalar (anti-) ghost fields, $(\phi_{2}) \phi_{1}$ are the massless
($ \Box \phi_{1} = \Box \phi_{2} = 0 $) (pseudo)scalar fields, $(\rho)\lambda $ are the {\it fermionic} 
($ \rho^{2} = \lambda^{2} = 0 ,\; \rho \,\lambda + \lambda \,\rho = 0,$ etc.) auxiliary ghost fields and $ B_{\mu\nu}$ is the Abelian 2-form 
[$B^{(2)} = (\frac {dx^\mu \wedge dx^\nu}{2!})\, B_{\mu\nu}$] gauge field.

We note that the above mentioned {\it off-shell nilpotent} ($s^2_{(a)d} = 0,\, s^2_{(a)b} = 0$) (anti-)co-BRST ($s_{(a)d}$) and (anti-)BRST ($s_{(a)b}$)
transformations (see, e.g. [22, 23] for details)
\begin{eqnarray}
&&s_{ad}\,B_{\mu\nu} = - \varepsilon_{\mu\nu\eta\kappa}\partial^{\eta}C^{\kappa}, \quad s_{ad}\,C_{\mu} = \partial_{\mu}\beta , 
\quad \quad  s_{ad}\,{\bar C}_{\mu} = {\bar{\cal B}}_\mu,\nonumber \\
&& s_{ad}\phi_{2} = \lambda , \quad \quad  s_{ad} \bar{\beta} = \rho ,\quad 
s_{ad}[\rho , \lambda , \beta , \phi_{1}, {\bar{\cal B}}_\mu,{\bar B}_\mu,\partial^{\nu} B_{\nu\mu}] = 0,
\end{eqnarray}
\begin{eqnarray}
&& s_{d}B_{\mu\nu} = - \varepsilon_{\mu\eta\nu\kappa}\partial^{\eta}\bar{C^{\kappa}},\quad \quad s_{d}\bar{C_{\mu}} 
= -\partial_{\mu}\bar{\beta},  \quad  \quad s_{d}C_{\mu} = - {\cal B_{\mu}}, \nonumber \\
&& s_{d}\phi_{2} = \rho , \quad \quad s_{d}\beta = - \lambda ,\quad \quad 
s_{d}[\rho ,\lambda ,\bar{\beta}, \phi_{1}, {\cal B_{\mu}}, B_{\mu},\partial^{\nu}B_{\nu\mu}] = 0,
\end{eqnarray}
\begin{eqnarray}
&& s_{ab}\,B_{\mu \nu} =  - (\partial_{\mu}\bar{C_{\nu}} - \partial_{\nu}\bar{C_{\mu}}) ,\quad \quad 
s_{ab}\, \bar C_{\mu} = - \partial_{\mu}\bar{\beta}, \quad \quad    s_{ab}\, C_\mu = {\bar B}_{\mu},\nonumber \\
&& s_{ab}\,\phi_{1} = - \rho , \quad \quad s_{ab}\,\beta = - \lambda ,  
\quad s_{ab}\,[\rho ,\lambda ,\bar{\beta} , \phi_{2} , {\bar{\cal B}}_\mu , \bar{B_{\mu}}, H_{\mu\nu\kappa}] = 0, 
\end{eqnarray}
\begin{eqnarray}
&& s_{b}\,B_{\mu\nu} = - (\partial_{\mu}C_{\nu} - \partial_{\nu}C_{\mu}),\quad \quad
s_{b}\, C_{\mu} = - \partial_{\mu}\beta ,\quad \quad s_{b}\, \bar C_{\mu} = - B_{\mu},  \nonumber \\
&& s_{b}\, \phi_{1} = - \lambda , \quad   \quad     s_{b}\,\bar{\beta} = -\rho ,\quad \quad  
s_{b}\,[\rho , \lambda , \beta , \phi_{2},{\cal B_{\mu}}, B_{\mu}, H_{\mu\nu\kappa}] = 0,
\end{eqnarray}
are continuous {\it symmetry} transformations of the action integrals ($ S_1 = \int d^{4}x$\\ ${\cal L}_{(B,\,\cal B)}$, 
$S_2 = \int d^{4}x \,{\cal L}_{(\bar{B},\,\bar{\cal{B}})}$) corresponding to the Lagrangian densities (1) and (2). Furthermore, 
the above transformations are {\it absolutely anticommuting} (i.e. $ s_{b}\,s_{ab} + s_{ab}\,s_{b} = 0$) in nature on the constrained 
hypersurface (in the 4$D$ Minkowskian background spacetime manifold) which is described by the following Curci-Ferrari (CF) 
type field equations\footnote{We are allowed to consider the CF-type restrictions as 
$ B_\mu - \bar B_\mu = \pm \,\alpha \,\partial_\mu \phi_1, {\cal B}_\mu - \bar{\cal B}_\mu  = \pm \,\beta\,\partial_\mu\phi_2$
where $\alpha$ and $\beta$ are numerical factors of any arbitrary value. However, the choices made in (7) are correct
because these are (anti-)BRST as well as (anti-)co-BRST invariant and consistent with absolute anticommutativity properties 
(cf. Eqns. (3)-(6) and (52)).}, namely;
\begin{eqnarray}
 {\cal{B}}_{\mu} - {\bar{\cal B}}_\mu = - \partial_{\mu}\phi_{2},\quad \qquad\qquad \;\;B_{\mu} - \bar{B_{\mu}} = - \partial_{\mu}\phi_{1},
 \end{eqnarray}
 which are found to be (anti-)co-BRST as well as (anti-)BRST invariant (i.e $ s_{(a)d}\, \big[ {\cal B}_{\mu} - \bar {\cal B}_{\mu} + \partial_{\mu}\phi_{2} \big] = 0$, $ s_{(a)b}\, \big[ B_{\mu} - {\bar B}_{\mu} + \partial_{\mu}\phi_{1} \big] = 0$) quantities. In fact,
both the above relations of (7) are the ones that are useful in proving the absolute anticommutativity property ($s_b\, s_{ab} + s_{ab}\, s_b = 0,\, 
s_d\, s_{ad} + s_{ad}\, s_d = 0$). 
 Furthermore, on the constrained hypersurface (defined by the field equations (7)), the action integrals $S_1$ and $S_2$ respect {\it all}
 the nilpotent symmetry transformations (see, e.g. [23] for details) listed in (3), (4), (5) and (6).


It can be explicitly checked that, under the nilpotent (anti-)co-BRST symmetry transformation $(s_{(a)d})$ 
and (anti-)BRST symmetry transformation $(s_{(a)b})$, the Lagrangian densities (1) and (2) transform as (see, e.g. [22, 23] for details): 
\begin{eqnarray}
s_{ad}\,{\cal L}_{(\bar{B},\bar{\cal{B})}}  =  \partial_{\mu}[(\partial^{\mu}C^{\nu} - \partial^{\nu}C^{\mu})\, {\bar{\cal B}}_\nu 
+ \rho\,\partial^{\mu}\beta + \lambda\, {\bar{\cal B}}^\mu ],
\end{eqnarray}
\begin{eqnarray}
s_{d}\,{\cal L}_{(B,\cal B)}  =  \partial_{\mu}[(\partial^{\mu}\bar C^{\nu} - \partial^{\nu}{\bar C}^\mu){\cal B}_{\nu} 
- \lambda\,\partial^{\mu} {\bar \beta} - \rho \,\cal B^{\mu}],
\end{eqnarray}
\begin{eqnarray}
 s_{ab} \,{\cal L}_{(\bar{B},\bar{\cal{B})}} = - \partial_{\mu}[(\partial^{\mu}\bar{C^{\nu}} - \partial^{\nu}\bar{C^{\mu}})\bar{B_{\nu}} 
- \rho\, \bar{B^{\mu}} + \lambda \,\partial^{\mu}\bar{\beta}],
\end{eqnarray}
\begin{eqnarray}
s_{b}\,{\cal L} _{(B,\cal B)} = - \partial_{\mu}[(\partial^{\mu}C^{\nu} - \partial^{\nu}C^{\mu})B_{\nu} + \rho\, \partial^{\mu}\beta + \lambda \,B^{\mu}].
\end{eqnarray}
We have not mentioned here the symmetry transformation properties of $ s_{ab}\,{\cal L} _{(B,\,{\cal B})}$,  
$ s_{b}\,{\cal L}_{(\bar{B},\,\bar{\cal{B})}}$,  $ s_{d}\,{\cal L}_{(\bar{B},\,\bar{\cal{B})}}$ and 
$ s_{ad}\,{\cal L} _{(B,\,{\cal B})}$ but it can be checked that, under these symmetry transformations, the Lagrangian 
densities transform to the total spacetime derivative {\it plus} the terms that are $\it zero$ on the constrained 
hypersurface defined by the CF-type restriction (7). Thus, we conclude that {\it both} the Lagrangian densities (1) and (2) 
respect the (anti-)BRST as well as (anti-)co-BRST symmetries {\it together} on the constrained hypersurface where the CF-type conditions (7) is satisfied (see, e.g. [23] for details).

\noindent
\section {Nilpotent (anti-)co-BRST Symmetries: Use of the Chiral and Anti-chiral Superfields and their Super-expansions} 

In this section, first of all, we derive the co-BRST (i.e. dual-BRST) transformations ($s_d$) by exploiting
the anti-chiral superfields which are the generalizations of the dynamical fields of our present theory onto 
the (4, 1)-dimensional anti-chiral super-submanifold as
\begin{eqnarray}
 B_{\mu\nu} (x) \longrightarrow {\tilde B}_{\mu\nu} (x, \bar\theta) &=& B_{\mu\nu}(x) + \bar\theta\, R_{\mu\nu} (x),\nonumber\\
 C_\mu (x) \longrightarrow {\tilde {\cal F}}_\mu (x, \bar\theta) &=& C_\mu(x) + \bar\theta\, B^{(1)}_\mu (x),\nonumber\\
 \bar C_\mu (x) \longrightarrow \tilde {\bar {\cal F}}_\mu (x, \bar\theta) &=& \bar C_\mu(x) + \bar\theta\, B^{(2)}_\mu (x),\nonumber\\
 \phi_1 (x)  \longrightarrow \tilde \Phi_1 (x, \bar\theta) &=& \phi_1 (x) + \bar\theta\, f_1 (x),\nonumber\\
 \phi_2 (x)  \longrightarrow \tilde \Phi_2 (x, \bar\theta) &=& \phi_2 (x) + \bar\theta\, f_2 (x), \nonumber\\
 \beta (x)  \longrightarrow \tilde \beta (x, \bar\theta) &=& \beta (x) + \bar\theta\, f_3 (x), \nonumber\\
 \bar\beta (x)  \longrightarrow \tilde {\bar\beta} (x, \bar\theta) &=& \bar\beta (x) + \bar\theta \,f_4 (x),
\end{eqnarray}
where the super-submanifold is parameterized by ($x^\mu, \bar\theta$). In the above, the {\it ordinary} fields 
($B_{\mu\nu} (x),\, C_\mu (x),\, \bar C_\mu (x),\,\phi_1 (x),\, \phi_2 (x),\, \beta (x),\, \bar\beta (x)$) 
are generalized onto this super-submanifold as 
(${\tilde B}_{\mu\nu} (x, \bar\theta),\,{\tilde {\cal F}}_\mu (x, \bar\theta),\, 
\tilde {\bar {\cal F}}_\mu (x, \bar\theta),\, \tilde \Phi_1 (x, \bar\theta), \tilde \Phi_2 (x, \bar\theta),\,
\tilde \beta (x, \bar\theta),\,\tilde {\bar\beta} (x, \bar\theta)$) where the seconadry fields ($R_{\mu\nu} (x),\; f_1 (x),\; f_2 (x),
f_3 (x),\, f_4 (x)$), on the r.h.s. of (12), are {\it fermionic} in nature and the pair ($ B^{(1)}_\mu (x),  B^{(2)}_\mu (x)$) 
are bosonic secondary fields. We have to determine these secondary fields in terms of the basic and 
auxiliary fields of our {\it ordinary} 4$D$ (anti-)co-BRST invariant theory described
by the Lagrangian densities (1) and (2) by invoking some appropriate (anti-)co-BRST invariant restrictions (CBRSTIRs). In this
connection, it is worthwhile to point out that we have {\it not} taken the generalizations {\it and} super-expansions
of the superfields corresponding to the auxiliary field $\rho,\, \lambda,\, B_\mu,\, {\bar B}_\mu,\, {\cal B}_\mu,\,{\bar{\cal B}}_\mu$
because these fields remain {\it invariant} under the (anti-)co-BRST symmetry transformations {\it and} they can be {\it accommodated} 
as the values of the secondary fields in (12).

We note that the total {\it gauge-fixing} term remains invariant under the (anti-)co-BRST symmetry transformations 
(i.e. $s_{a(d)}\, [\partial^\nu B_{\nu\mu} (x)] = 0$ and $s_{a(d)}\, \phi_1(x) = 0$). As a consequence, first of all, 
we demand that the anti-chiral superfield $\tilde \Phi_1 (x,\,\bar\theta)$ should be independent
of the ``soul" coordinate $\bar\theta$. Thus, we have the following restriction on this superfield, namely;
\begin{eqnarray}
\tilde \Phi_1 (x, \bar\theta) = \phi_1 (x) \qquad \Longrightarrow \qquad f_1 (x) = 0.
\end{eqnarray}
For the sake of generality of our present approach, we do {\it not} utilize the restriction
$s_{(a)d}\,(\partial^\nu B_{\nu\mu}(x)) = 0$ because this is connected with the dual-HC where the differential geometric operator
$\delta = -\,*\,d\,*$ plays an important role because $\delta B^{(2)} = (\partial^\nu B_{\nu\mu}) dx^\mu$. 
Furthermore, it can be readily verified that the following very {\it useful} set of quantities are found to 
be co-BRST invariant, namely;
\begin{eqnarray}
&& s_d\, \bar\beta = 0,\qquad \quad s_d\, ( \lambda\,\beta) = 0, \quad\qquad s_d (\rho\,\phi_2) = 0, \nonumber\\
&& s_d \Big[ C_\mu\, \partial^\mu \rho +  {\cal B}_\mu\, \partial^\mu \phi_2 \Big] = 0, \quad
s_d \Big[ {\bar C}_\mu\, \partial^\mu \lambda - \partial_\mu \bar\beta\,\partial^\mu \beta \Big] = 0, \nonumber\\
&& s_d \Big[ \frac{1}{2}\, \varepsilon_{\mu\nu\eta\kappa}\, \partial^\mu {\cal B}^\nu\, B^{\eta\kappa} - 
\Big( \partial_\mu {\bar C}_\nu - \partial_\nu {\bar C}_\mu \Big) \Big(\partial^\mu C^\nu \Big) \Big] = 0.
\end{eqnarray}
The above claim can be checked explicitly by exploiting the co-BRST symmetry transformations ($s_d$) given in (4). As a
consequence, the above set of invariant quantities, generalized onto the (4, 1)-dimensional anti-chiral super-submanifold,
would remain independent of the ``soul" coordinate $\bar\theta$. 
This requirement is one of the key features of the augmented version of (anti-)chiral
superfield formalism. In other words, we have the following co-BRST invariant restrictions on the anti-chiral superfields:  
\begin{eqnarray}
&&\tilde {\bar\beta} (x, \bar\theta) = \bar\beta (x), \quad \;  \lambda (x)\,\tilde \beta (x, \bar\theta) = \lambda (x)\, \beta (x),
\quad \; \rho (x)\, \tilde \Phi_2 (x, \bar\theta)\, = \rho (x)\,\phi_2 (x), \nonumber\\
&&\tilde{\cal F}_\mu (x,\bar\theta) \partial^\mu \rho (x) +  {\cal B}_\mu\, \partial^\mu \tilde \Phi_2 (x, \bar\theta) = 
C_\mu (x)\,\partial^\mu \rho (x) +  {\cal B}_\mu\, \partial^\mu\phi_2(x), \nonumber\\
&& \tilde {\bar{\cal F}}_\mu (x,\bar\theta) \partial^\mu \lambda (x) 
- \partial_\mu \tilde {\bar\beta} (x, \bar\theta) \partial^\mu \tilde \beta (x, \bar\theta)\, = 
{\bar C}_\mu(x)\,\partial^\mu  \lambda (x) -  \partial_\mu \bar\beta (x)\,\partial^\mu \beta (x),\nonumber\\ 
&&\frac{1}{2}\, \varepsilon_{\mu\nu\eta\kappa} \partial^\mu {\cal B}^\nu (x)\, {\tilde B}^{\eta\kappa} (x, \bar\theta)
- \Big( \partial_\mu \tilde {\bar {\cal F}}_\nu (x, \bar\theta) - \partial_\nu \tilde {\bar{\cal F}}_\mu (x, \bar\theta)\Big) 
\partial^\mu \tilde {\cal F}^\nu (x, \bar\theta) \nonumber\\
&& = \frac{1}{2}\, \varepsilon_{\mu\nu\eta\kappa} \partial^\mu {\cal B}^\nu (x)\, B^{\eta\kappa} (x)
- \Big( \partial_\mu \bar C_\nu (x) - \partial_\nu \bar C_\mu (x)\Big) \partial^\mu C^\nu (x).
\end{eqnarray}
We note that the fields on the r.h.s. of the above equality are {\it only} function of the ordinary spacetime coordinates ($x^\mu$)
but the (super)fields on the l.h.s. are function of the anti-chiral superspace coordinates ($x^\mu, \bar\theta$). Thus,  after substitution of
the superfields (12) into the expressions of the l.h.s., we have to set the coefficients of $\bar\theta$-variable equal to zero. This exercise yields
the following relationships amongst the {\it secondary} fields and the {\it basic} as well as {\it auxiliary} fields of the ordinary 4$D$ gauge theory:
\begin{eqnarray}
&& R_{\mu\nu} (x) = - \varepsilon_{\mu\nu\eta\kappa}\, \partial^\eta \bar C^\kappa (x), \quad B^{(2)}_\mu (x) = - \partial_\mu \bar\beta (x), \quad
f_2 (x) =  \rho (x), \nonumber\\  && B^{(1)}_\mu (x) = - {\cal B}_\mu (x), \qquad f_4 (x) = 0, \qquad f_3(x) = - \lambda (x).
\end{eqnarray}
The substitution of these values into the expansions (12) yields
\begin{eqnarray}
&& {\tilde B}^{(d)}_{\mu\nu} (x, \bar\theta) = B_{\mu\nu}(x) + \bar\theta\, \Big(- \varepsilon_{\mu\nu\eta\kappa}\, \partial^\eta \bar C^\kappa (x) \Big)
\equiv B_{\mu\nu}(x) + \bar\theta\,\Big( s_d\, B_{\mu\nu}(x) \Big),\nonumber\\
&& {\tilde {\cal F}}^{(d)}_\mu (x, \bar\theta) = C_\mu(x) + \bar\theta\, \Big(- {\cal B}_\mu (x) \Big) \equiv C_\mu(x) + \bar\theta\,\Big( s_d\,C_\mu (x) \Big),\nonumber\\
&& \tilde {\bar {\cal F}}^{(d)}_\mu (x, \bar\theta) = \bar C_\mu(x) + \bar\theta\,\Big( - \partial_\mu \bar\beta (x) \Big) \equiv 
\bar C_\mu(x) + \bar\theta\,\Big( s_d\, \bar C_\mu (x) \Big),\nonumber\\
&&  \tilde \Phi^{(d)}_1 (x, \bar\theta) = \phi_1 (x) +\bar\theta\, \Big( 0 \Big) \equiv  \phi_1 (x) + \bar\theta\, \Big( s_d \phi_1 (x) \Big),\nonumber\\
&& \tilde \Phi^{(d)}_2 (x, \bar\theta) = \phi_2 (x) + \bar\theta \, \Big(  \rho (x) \Big) \equiv \phi_2 (x) + \bar\theta \, \Big(s_d\, \phi_2 (x) \Big), \nonumber\\
&& \tilde \beta^{(d)} (x, \bar\theta) = \beta (x) + \bar\theta \,\Big( - \lambda (x) \Big) \equiv \beta (x) + \bar\theta \,\Big( s_d\, \beta (x) \Big), \nonumber\\
&& \tilde {\bar\beta}^{(d)} (x, \bar\theta) = \bar\beta (x) + \bar\theta\, \Big(0 \Big) \equiv \bar\beta (x) + \bar\theta\, \Big(s_d\, \bar\beta \Big),
\end{eqnarray}
where the superscript ($d$) on the superfields denotes the expansions of the superfields after the application of dual-BRST (i.e. co-BRST)
invariant restrictions [cf. (13), (15)] on the anti-chiral (super)fields of our present theory. 
A close look at the above equation demonstrates that
we have already obtained the dual-BRST (i.e. co-BRST) symmetry transformations (4) of our theory.
These are nothing but the coefficients of the Grassmannian variable $\bar\theta$ in the super expansions of the superfields.

We would like to remark on the relationships (16) that have been obtained amongst the 
secondary fields and the basic/auxiliary fields of our present theory. The co-BRST
invariant restrictions (15) produce, actually, the relationships
$f_3(x)\,\lambda(x) = 0$ and $f_2(x)\,\rho(x) = 0$. Thus, we have the freedom to choose
$f_2(x)$ and $f_3(x)$ proportional to $\rho(x)$ and $\lambda(x)$, respectively. We have taken this liberty to choose $f_2(x) = \rho(x)$
and $f_3(x) = -\,\lambda(x)$ which are very much essential because these choices satisfy the restrictions that have been written in the 
{\it second} and {\it third} rows of Eqn. (15). 
With these choices, we obtain all the {\it exact} relationships that have been quoted in Eqn. (16). In fact, it turns out that
the choice $f_2(x) = \rho(x)$ produces the result $B_\mu^{(1)} = -\,{\mathcal B}_\mu (x)$. In exactly similar
fashion, the choice $f_3(x) = -\,\lambda(x)$ leads to the derivation of the bosonic secondary field ($B_\mu ^{(2)}$) 
as: $B_\mu ^{(2)} = -\,\partial_\mu\,\bar \beta(x)$.

To derive the proper anti-co-BRST symmetry transformations, we invoke here the chiral superfields, 
parameterized by the super-coordinates ($x^\mu, \,\theta$), as the generalizations of the ordinary fields 
in the following fashion:
\begin{eqnarray}
&& B_{\mu\nu} (x) \longrightarrow {\tilde B}_{\mu\nu} (x, \theta) = B_{\mu\nu}(x) + \theta\, \bar R_{\mu\nu} (x),\nonumber\\
&& C_\mu (x) \longrightarrow {\tilde {\cal F}}_\mu (x, \theta) = C_\mu(x) + \theta\, \bar B^{(1)}_\mu (x),\nonumber\\
&& \bar C_\mu (x) \longrightarrow \tilde {\bar {\cal F}}_\mu (x, \theta) = \bar C_\mu(x) + \theta\, \bar B^{(2)}_\mu (x),\nonumber\\
&& \phi_1 (x)  \longrightarrow \tilde \Phi_1 (x, \theta) = \phi_1 (x) + \theta \bar f_1 (x),\nonumber\\
&& \phi_2 (x)  \longrightarrow \tilde \Phi_2 (x, \theta) = \phi_2 (x) + \theta \bar f_2 (x), \nonumber\\
&& \beta (x)  \longrightarrow \tilde \beta (x, \theta) = \beta (x) + \theta \bar f_3 (x), \nonumber\\
&& \bar\beta (x)  \longrightarrow \tilde {\bar\beta} (x, \theta) = \bar\beta (x) + \theta \bar f_4 (x),
\end{eqnarray}
where the 4$D$ fields ($B_{\mu\nu} (x),\, \phi_1 (x),\, \phi_2 (x),\, \beta (x),\, \bar\beta (x),\,\bar B^{(1)}_\mu (x),\, \bar B^{(2)}_\mu (x) $)
are bosonic and ($\bar R_{\mu\nu} (x),\, C_\mu (x),\, \bar C_\mu (x),\, \bar f_1 (x),\, \bar f_2 (x),\, \bar f_3 (x),\, \bar f_4 (x) $) are fermionic 
due to the ``fermionic" nature of the Grassmannian variable $\theta$. 
Out of these fields, we note that the {\it secondary} fields ($\bar R_{\mu\nu} (x),\, \bar f_1 (x),\, \bar f_2 (x),\,
 \bar f_3 (x),\, \bar f_4 (x),$\\ ${\bar B}^{(1)}_\mu (x),\, {\bar B}^{(2)}_\mu (x)$) are to be determined by exploiting the virtues 
of the  anti-co-BRST invariant restrictions on the {\it chiral} superfields (defined on our chosen chiral super-submanifold). 
Towards this goal in mind, we collect below some of the {\it useful} anti-co-BRST invariant
quantities that are present in our theory.

We observe that the following quantities are anti-co-BRST invariant:
\begin{eqnarray}
&& s_{ad}\,\beta = 0,\qquad \quad s_{ad}\, (\rho\, \bar\beta) = 0, \quad\qquad s_{ad} (\lambda\, \phi_2) = 0, \nonumber\\
&& s_{ad} \Big[ {\bar C}_\mu\, \partial^\mu \lambda - {\bar {\cal B}}_\mu\, \partial^\mu \phi_2 \Big] = 0, \quad
s_{ad} \Big[  C_\mu\, \partial^\mu \rho - \partial_\mu \bar\beta\,\partial^\mu \beta \Big] = 0, \nonumber\\
&& s_{ad} \Big[ \frac{1}{2}\, \varepsilon_{\mu\nu\eta\kappa}\, \partial^\mu {\bar{\cal B}}^\nu\, B^{\eta\kappa} -
\Big(\partial_\mu \bar C_\nu \Big)\Big( \partial^\mu  C^\nu - \partial^\nu C^\mu \Big)  \Big) \Big]= 0.
\end{eqnarray}
As a consequence, the above quantities are physical because, in a field theoretic model for the Hodge theory, the (anti-)BRST
as well as the (anti-)co-BRST invariant quantities {\it ought} to be ``physical". Thus, their generalizations on the (4, 1)-dimensional 
chiral super-submanifold must be independent of the Grassmannian variable $\theta$. Based on this argument, 
we have the following restrictions on the chiral (super)fields, namely;
\begin{eqnarray}
&&\tilde \beta (x, \theta) = \beta (x), \quad \; \rho (x)\, \tilde {\bar\beta }(x, \theta) = \rho (x)\, \bar\beta (x),
\quad \; \lambda (x)\, \tilde \Phi_2 (x, \theta) = \lambda (x)\, \phi_2 (x), \nonumber\\
&&\tilde{\cal F}_\mu (x,\theta) \partial^\mu \lambda (x) -  {\bar{\cal B}}_\mu (x)\, \partial^\mu \tilde \Phi_2 (x, \theta) 
= C_\mu (x)\,\partial^\mu \lambda (x) - {\bar{\cal B}}_\mu (x)\, \partial^\mu\phi_2(x), \nonumber\\
&& \tilde {\bar{\cal F}}_\mu (x,\theta) \partial^\mu \rho (x) 
- \partial_\mu \tilde {\bar \beta} (x, \theta)\, \partial^\mu \tilde \beta (x, \theta) = 
{\bar C}_\mu (x)\,\partial^\mu  \rho (x) - \partial_\mu \bar\beta (x)\, \partial^\mu \beta (x),\nonumber\\ 
&&\frac{1}{2}\, \varepsilon_{\mu\nu\eta\kappa} \partial^\mu {\bar{\cal B}}^\nu (x)\, {\tilde B}^{\eta\kappa} (x, \theta)
- \partial_\mu \tilde {\bar{\cal F}}_\nu (x, \theta) \Big( \partial^\mu \tilde {\cal F}^\nu (x, \theta)
- \partial^\nu \tilde {\cal F}^\mu(x, \theta)\Big) \nonumber\\
&& = \frac{1}{2}\, \varepsilon_{\mu\nu\eta\kappa} \partial^\mu {\bar{\cal B}}^\nu (x)\, B^{\eta\kappa} (x)
- \partial_\mu \bar C_\nu (x)\Big( \partial^\mu C^\nu (x) - \partial^\nu C^\mu (x)\Big).
\end{eqnarray}
It will be noted that we have {\it not} taken the super-expansions of the superfields corresponding to the 
auxiliary fields $\rho (x),\, \lambda(x)$, etc., because these are (anti-) co-BRST invariant quantities 
(i.e. $s_{(a)d}\, \rho = 0,\, s_{(a)d}\, \lambda = 0$).

The above restrictions in (20) lead to the derivation of the {\it secondary} fields, in terms of the {\it basic}
and {\it auxiliary} fields of the ordinary 4$D$ theory, as
\begin{eqnarray}
&& \bar R_{\mu\nu} (x) = - \varepsilon_{\mu\nu\eta\kappa}\, \partial^\eta  C^\kappa (x), \quad \bar B^{(1)}_\mu (x) =  \partial_\mu \beta (x), \quad
\bar f_2 (x) =  \lambda (x), \nonumber\\  && \bar B^{(2)}_\mu (x) =  {\bar{\cal B}}_\mu (x),\qquad \bar f_3(x) = 0, \qquad \bar f_4 (x) = \rho (x).
\end{eqnarray}
The substitution of the above secondary fields into the chiral super-expansions (18) 
leads to the following explicit expansions of the chiral superfields, namely;   
\begin{eqnarray}
&& {\tilde B}^{(ad)}_{\mu\nu} (x, \theta) = B_{\mu\nu}(x) + \theta\, \Big(- \varepsilon_{\mu\nu\eta\kappa}\, \partial^\eta C^\kappa (x) \Big)
\equiv B_{\mu\nu}(x) + \theta\,\Big( s_{ad}\, B_{\mu\nu}(x) \Big),\nonumber\\
&& {\tilde {\cal F}}^{(ad)}_\mu (x, \theta) = C_\mu(x) + \theta\, \Big( \partial_\mu \beta (x) \Big) \equiv C_\mu(x) + \theta\,\Big( s_{ad}\, C_\mu (x) \Big),\nonumber\\
&& \tilde {\bar {\cal F}}^{(ad)}_\mu (x, \theta) = \bar C_\mu(x) + \theta\,\Big( {\bar {\cal B}}_\mu (x) \Big) \equiv 
\bar C_\mu(x) + \theta\,\Big( s_{ad}\, \bar C_\mu (x) \Big),\nonumber\\
&&  \tilde \Phi^{(ad)}_1 (x, \theta) = \phi_1 (x) + \theta\, \Big( 0 \Big) \equiv  \phi_1 (x) + \theta\, \Big( s_{ad} \phi_1 (x) \Big),\nonumber\\
&& \tilde \Phi^{(ad)}_2 (x, \theta) = \phi_2 (x) + \theta \, \Big( \lambda (x) \Big) \equiv \phi_2 (x) + \theta \, \Big( s_{ad}\, \phi_2 (x) \Big), \nonumber\\
&& \tilde \beta^{(ad)} (x, \theta) = \beta (x) + \theta \,\Big( 0 \Big) \equiv \beta (x) + \theta \,\Big( s_{ad}\, \beta (x) \Big), \nonumber\\
&& \tilde {\bar\beta}^{(ad)} (x, \theta) = \bar\beta (x) + \theta\, \Big( \rho (x) \Big) \equiv \bar\beta (x) + \theta\, \Big(s_{ad}\, \bar\beta (x)\Big),
\end{eqnarray}
where the superscript $(ad)$ denotes the superfields that have been obtained after the application of the anti-CBRSTIRs (20) on the superfields.
We note that we have already obtained the anti-co-BRST symmetry transformations of our present theory (cf. (3)).
These transformations appear as the coefficients of the variable $\theta$ on the r.h.s. of (22)
(i.e. the chiral expansions of the superfields).

We remark here on the derivation of the relationships amongst the secondary fields of the expansions (18) and 
basic/auxiliary fields of the starting Lagrangian densities (1). In this connection, it is worthwhile to point out 
that we obtain the relationships $\rho(x)\, \bar f_4 (x) = 0$ and $\lambda(x) \bar f_2(x) = 0$ which show that 
$\bar f_4 (x) \propto \rho(x)$ and $\bar f_2 (x) \propto \lambda(x)$ as non-trivial solutions from the top row of
restrictions (20). However, the second and third rows of the anti-co-BRST invariant 
restrictions (20) are satisfied  if and only if we choose $\bar f_2 (x) = \lambda(x), \, \bar f_4(x) = \rho(x),\, 
{\bar B}^{(2)}_\mu(x) = {\bar {\cal B}}_\mu (x)$ and ${\bar B}^{(1)}_\mu (x)= \partial_\mu \beta (x)$. In the last entry of the restrictions (20), we have to take the help of {\it all} the above quoted relationships (e.g. $\bar f_2 (x) = \lambda (x),\, \bar f_4 (x) = \rho(x),\, {\bar B}^{(2)}_\mu(x) = {\bar {\cal B}}_\mu(x),\, {\bar B}^{(1)}_\mu(x) = \partial_\mu \beta (x)$) to obtain $R_{\mu\nu} (x) = - \varepsilon_{\mu\nu\eta\kappa} \partial^\eta C^\kappa (x)$. Thus, we note that {\it all} the secondary fields of super expansions (18) are precisely determined in terms of the basic/auxiliary fields of the Lagrangian densities (1) which, ultimately, lead to the derivation  of the anti-co-BRST symmetries (cf. (3)) in the {\it final} super expansions listed in Eqn. (22)
as the coefficients of $\theta$.

From the (anti-)chiral super expansions (22) and (17), we obtain the geometrical meaning of the (anti-)co-BRST 
symmetry transformations due to the following relationships
\begin{eqnarray}
\frac{\partial}{\partial \theta}\, {\tilde \Omega}^{(ad)} (x, \theta) = s_{ad}\, \omega(x),\qquad \quad 
\frac{\partial}{\partial \bar\theta}\, {\tilde \Omega}^{(d)} (x, \bar\theta) = s_d\, \omega(x),
\end{eqnarray}
where ${\tilde \Omega}^{(ad)} (x, \theta)$ and ${\tilde \Omega}^{(d)} (x, \bar\theta)$ are the generic chiral and anti-chiral superfields 
obtained after the application of anti-CBRSTIRs and CBRSTIRs on the superfields and $\omega(x)$ is the generic 4$D$ ordinary bosonic
and fermionic fields of the ordinary 4$D$ (anti-)BRST and (anti-)co-BRST invariant theory. 
We continue to use the partial derivatives $\partial_\theta$ and
$\partial_{\bar\theta}$ because super-submanifolds are a part of general supermanifold  that is characterized by the superspace variable 
$Z^M = (x^\mu,\, \theta,\, \bar\theta)$ and (anti-)chiral superfields are limiting
cases of the general superfields where we set $\theta = 0$ and/or $\bar\theta = 0$, respectively.
Geometrically, the above equation implies that the (anti-)co-BRST symmetry transformations on the 4$D$ generic field $\omega(x)$ 
is equivalent to the translation of the (4, 2)-dimensional superfield (${\tilde \Omega}^{(ad)} $)${\tilde \Omega}^{(d)}$
 along the $(\theta)\bar\theta$-directions of the (4, 1)-dimensional chiral and anti-chiral super-submanifolds (of the 
general (4, 2)-dimensional supermanifold on which our 4$D$ theory is generalized). Thus, the nilpotency property of $s_{(a)d}$ is 
intimately connected with the nilpotency property of $(\partial_\theta)\partial_{\bar\theta}$
(i.e. $s^2_{(a)d} = 0 \Leftrightarrow (\partial_\theta)^2 = (\partial_{\bar\theta})^2 = 0$).

\noindent
\section {Nilpotent (anti-)BRST Symmetries: Use of the Chiral and Anti-chiral Superfields and their Super-expansions} 

In our present section, we derive the nilpotent (anti-)BRST symmetry transformations by using the 
augmented (anti-)chiral superfield approach. First of all, we concentrate on the generalization of 4$D$ dynamical fields of Lagrangian density (1) to
 the anti-chiral superfields which are parameterized by the variables  ($x^\mu, \bar\theta$)
(cf. Eqn. (12)). It can be seen that the kinetic term remains invariant (i.e. $s_{(a)b}\, H_{\mu\nu\eta} = 0,\, s_{(a)b}\, \phi_2 = 0$) 
under the BRST as well as anti-BRST symmetry transformations. Hence, the anti-chiral superfield
$\tilde {\Phi}_2 (x, \bar\theta)$ should be unchanged.
In other words, the superfield $\tilde{\Phi}_2 (x, \bar\theta)$ must be 
independent of the Grassmannian variable $\bar\theta$ due to the basic requirements of augmented (anti-)chiral superfield formalism. 
In fact, this requirement leads us to:
\begin{eqnarray}
\tilde {\Phi}_2 (x, \bar\theta) = \phi_2 (x) \qquad \quad \Longrightarrow \qquad \quad  f_2 (x) = 0.
\end{eqnarray}
It should be noted that we have {\it not} taken into account $s_{(a)b}\,H_{\mu\nu\eta} = 0$ because
it is connected with the HC (owing its origin to $d = dx^\mu\partial_\mu$) which has been used in our earlier work [17].
To utilize the BRSTIRs, we have to find out a set of {\it useful} BRST invariant quantities. In this connection, 
we see that if we apply the BRST symmetry transformations on the following quantities, they turn out to be zero, namely;
\begin{eqnarray}
&& s_b\,{\beta}  =  0, \qquad s_b\,\big( \phi_{1}\, \lambda\big)  =  0,\qquad s_b\,\big( \bar\beta \, \rho\big) = 0, \nonumber\\ 
&& s_b\,\big[\,\bar C_{\mu}\partial^{\mu}\lambda - B_{\mu}\,\partial^{\mu}\phi_{1}\big] = 0,
\qquad s_b\,\big[C_\mu\,\partial^\mu\rho - \partial_\mu \bar{\beta}\, \partial^\mu\beta \big] = 0, \nonumber\\
&&s_b\,\big[ (\partial^\mu  B^\nu)\,B_{\mu\nu} - \partial^\mu {\bar C}^\nu \,(\partial_\mu C_\nu - \partial_\nu C_\mu)\,\big] = 0. 
\end{eqnarray}
Due to the invariance under the BRST symmetry transformations, the above quantities would be unaffected by the 
``soul" coordinate $\bar\theta$ when these quantities are generalized onto the (4, 1)-dimensional anti-chiral 
super-submanifold. This is in accordance with the basic tenets of augmented (anti-)chiral superfield approach to BRST formalism. 
In other words, we have the following restrictions on the (super)fields, namely; 
\begin{eqnarray}
&&\tilde \beta (x, \bar\theta) = \beta (x), \quad \; \rho (x)\, \tilde {\bar\beta }(x, \bar\theta) = \rho (x)\, \bar\beta (x),
\quad \; \lambda (x)\, \tilde \Phi_1 (x, \bar\theta) = \lambda (x)\, \phi_1 (x), \nonumber\\
&&\tilde{\cal F}^\mu (x,\bar\theta)\; \partial_\mu \lambda (x) -  B^\mu (x)\; \partial_\mu \tilde \Phi_2 (x, \bar\theta) = 
C^\mu (x)\;\partial_\mu \lambda (x) - B^\mu (x)\, \partial_\mu\phi_2(x), \nonumber\\
&& \tilde {\bar{\cal F}}^\mu (x,\bar\theta)\; \partial_\mu \rho (x) 
- \partial^\mu \tilde {\bar \beta} (x, \bar\theta)\, \partial_\mu \tilde \beta (x, \bar\theta) = 
{\bar C}^\mu (x)\,\partial_\mu  \rho (x) - \partial^\mu \bar\beta (x)\, \partial_\mu \beta (x),\nonumber\\ 
&&\frac{1}{2}\, \varepsilon_{\mu\nu\eta\kappa} \partial^\mu  B^\nu (x)\, {\tilde B}^{\eta\kappa} (x, \bar\theta)
-  \partial^\mu \tilde {\bar{\cal F}}^\nu (x, \bar\theta) \Big( \partial_\mu \tilde {\cal F}_\nu (x, \bar\theta)- \partial_\nu \tilde {\cal F}_\mu(x, \bar\theta)\Big) \nonumber\\
&& = \frac{1}{2}\, \varepsilon_{\mu\nu\eta\kappa} \partial^\mu  B^\nu (x)\, B^{\eta\kappa} (x)
- \partial^\mu \bar C^\nu (x)\;\Big( \partial_\mu C_\nu (x) - \partial_\nu C_\mu (x)\Big).
\end{eqnarray}
Taking the super expansions (12) and substituting them into the above equation, we get the {\it secondary} 
fields as:
\begin{eqnarray}
&& R_{\mu\nu} (x) = -  (\partial_\mu C_\nu - \partial_\nu C_\mu )(x), \quad B^{(1)}_\mu (x) = - \partial_\mu \beta (x), \quad
f_1 (x) = - \lambda (x), \nonumber\\  && f_2 (x) = 0, \quad  B^{(2)}_\mu (x) = -  B_\mu (x), \quad f_4 (x) = - \rho (x), \quad f_3(x) = 0.
\end{eqnarray}
Putting these values into the  super-expansions (12), we get the following\footnote{There are {\it four} fermionic symmetries
in our theory. In general, we could have taken {\it four} Grassmannian variables ($\theta_1,\, \bar\theta_1,\, \theta_2,\, \bar\theta_2$)
to discuss the (anti-)co-BRST and (anti-)BRST symmetries where the pairs ($\theta_1,\,\bar\theta_1$) and ($\theta_2,\,\bar\theta_2$) 
could have been utilized {\it at a time} separately and independently. However, for the sake of brevity, we have taken only the pair ($\theta,\,\bar\theta$)
for our {\it whole} discussion on the fermionic symmetries and corresponding charges.}:
\begin{eqnarray}
&& {\tilde B}^{(b)}_{\mu\nu} (x, \bar\theta) = B_{\mu\nu}(x) + \bar\theta\, \Big(-  (\partial_\mu C_\nu - \partial_\nu C_\mu )(x) \Big)
\equiv B_{\mu\nu}(x) + \bar\theta\,\Big( s_b\, B_{\mu\nu}(x) \Big),\nonumber\\
&& {\tilde {\cal F}}^{(b)}_\mu (x, \bar\theta) = C_\mu(x) + \bar\theta\, \Big(- \partial_\mu \beta (x)\Big) \equiv C_\mu(x) + \bar\theta\,\Big( s_d\,C_\mu (x) \Big),\nonumber\\
&& \tilde {\bar {\cal F}}^{(b)}_\mu (x, \bar\theta) = \bar C_\mu(x) + \bar\theta\,\Big( -  B_\mu (x) \Big) \equiv 
\bar C_\mu(x) + \bar\theta\,\Big( s_b\, \bar C_\mu (x) \Big),\nonumber\\
&&  \tilde \Phi^{(b)}_1 (x, \bar\theta) = \phi_1 (x) +\bar\theta\, \Big(- \lambda (x) \Big) \equiv  \phi_1 (x) + \bar\theta\, \Big( s_b \phi_1 (x) \Big),\nonumber\\
&& \tilde \Phi^{(b)}_2 (x, \bar\theta) = \phi_2 (x) + \bar\theta \, \Big( 0 \Big) \equiv \phi_2 (x) + \bar\theta \, \Big(s_b\, \phi_2 (x) \Big), \nonumber\\
&& \tilde \beta^{(b)} (x, \bar\theta) = \beta (x) + \bar\theta \,\Big( 0 \Big) \equiv \beta (x) + \bar\theta \,\Big( s_b\, \beta (x) \Big), \nonumber\\
&& \tilde {\bar\beta}^{(b)} (x, \bar\theta) = \bar\beta (x) + \bar\theta\, \Big(- \rho (x) \Big) \equiv \bar\beta (x) + \bar\theta\, \Big(s_b\, \bar\beta \Big),
\end{eqnarray}
where the superscript ($b$), in the above equation on the anti-chiral superfields, denotes the anti-chiral super expansions 
after the application of BRST invariant restrictions. A careful and close look at (28) demonstrates that we have already
derived some of the off-shell nilpotent ($s^2_b = 0$) BRST symmetry transformations of (6) which are nothing but the
coefficient of the Grassmannian variable $\bar\theta$. This shows that we have: $s_b \leftrightarrow \partial_{\bar\theta}$.

A few remarks are in order as far as the derivation of relationships between secondary fields of expansions (12)
and basic/auxiliary fields of Lagrangian densities (1) are concerned. From the top three restrictions
in Eqn. (26), we obtain $f_3(x) = 0$, $f_1(x)\;\lambda(x) = 0$ and $f_4(x)\;\rho(x) = 0$. The latter two
conditions imply that the non-trivial solution is $f_1(x)\,\propto \,\lambda(x)$ and $f_4(x)\,\propto \,\rho(x)$. The
restrictions of second and third rows of (26) are satisfied if and only if we take into account
the specific secondary fields as:
$f_1(x) = -\lambda, \; f_4(x) = -\,\rho,\;B_\mu(x) = -\,B_\mu$ and $B_\mu^{(1)} = -\,\partial_\mu\,\beta$.
These inputs imply that we have already derived the expansions $\tilde {\cal F}^{(b)}_\mu(x,\bar\theta)$ and 
$\tilde {\bar{\cal F}}^{(b)}_\mu(x,\bar\theta)$. These expansions are now utilized in the last restrictions of Eqn. (26)
which leads to the derivation of $R_{\mu\nu}(x)$. Thus, we observe that {\it all} the secondary fields of expansion (12)
have been determined in terms of basic/auxiliary fields which, in turn, imply that we have derived {\it all}
the BRST symmetry transformations (cf. (28) and (6)) of our present 4$D$  gauge theory.

In order to derive the anti-BRST symmetry transformations, we have to use the chiral superfield mentioned in (18).  
Furthermore, it can be checked explicitly that the following quantities remain invariant under the anti-BRST symmetry transformations 
($s_{ab}$):
\begin{eqnarray}
&& s_{ab}\,\bar{\beta}  =  0, \qquad s_{ab}\,\big( \rho\, \phi_{1}\big)  =  0,\qquad s_{ab}\,\big( \lambda \, \beta \big) = 0, \nonumber\\ 
&& s_{ab}\,\big[\,C^{\mu}\partial_{\mu}{\rho} + \bar{B}^{\mu}\partial_{\mu}\phi_{1}\big] = 0,\qquad s_{ab}\,\big[\,\bar{C}^{\mu}\,\partial_{\mu}\lambda - \partial^{\mu}\bar{\beta}\,\partial_{\mu}\beta\,\big] = 0, \nonumber\\
&&s_{ab}\,\big[\,(\partial^{\mu}\bar{B}^{\nu})B_{\mu\nu} - (\partial^{\mu}\bar{C}^{\nu} - \partial^{\nu}\bar{C}^{\mu})\,\partial_{\mu}C_{\nu}\,\big] = 0. 
\end{eqnarray}
The above invariant quantities should remain independent of the ``soul" coordinate $\theta$  when they are generalized 
onto the (4, 1)-dimensional chiral super-submanifold. This statement can be mathematically expressed as:
\begin{eqnarray}
&& \tilde{\bar\beta} (x, \theta) = \bar\beta (x), \quad \lambda(x)\, \tilde{\beta} (x, \theta) =  \lambda (x) \, \beta (x),\quad \rho(x)\, {\tilde \Phi}_1 (x, \theta) = \rho (x)\, \phi_1(x), \nonumber\\
&& \tilde{\bar{\cal{F}}}^\mu (x, \theta)\,\partial_{\mu}\lambda (x) - \partial^{\mu}\tilde{\bar{\beta}} (x, \theta)\,\partial_{\mu}\tilde{\beta} (x, \theta) =  \bar{C}^{\mu} (x)\partial_{\mu}\,\lambda (x) - \partial^{\mu}\bar{\beta} (x)\,\partial_{\mu}\beta (x), \nonumber\\
&&\tilde{\cal F}^{\mu} (x, \theta)\,\partial_{\mu}\rho (x) + \bar{B}^{\mu}(x)\, \partial_{\mu}\tilde{\Phi}_{1} (x, \theta) 
=  C^{\mu} (x)\,\partial_{\mu}\rho (x) + \bar{B}^{\mu} (x)\,\partial_{\mu}\phi_{1} (x),\nonumber\\
&&\partial^{\mu}{\bar B}^\nu(x)\,\tilde{B}_{\mu\nu} (x, \theta) - \Bigl( \partial^{\mu} {\tilde{\bar {\cal F}}}^{\nu}(x, \theta) 
- \partial^{\nu} {\tilde{\bar {\cal F}}}^{\mu}(x, \theta) \Bigr)\,\partial_{\mu} {\tilde{\cal F}}_{\nu}(x, \theta) \nonumber\\ 
&& = \partial^{\mu} {\bar B}^{\nu}(x)\,B_{\mu\nu}(x) - \Bigl( \partial^{\mu}\bar{C}^{\nu}(x) - \partial^{\nu}\bar{C}^{\mu}(x) \Bigr)\,\partial_{\mu}\bar{C}_{\nu}(x).
\end{eqnarray}
After substituting the super expansions (18)  in the above equation, we get the {\it secondary} fields 
in terms of the {\it basic} as well as {\it auxiliary} fields as follows:
\begin{eqnarray}
&& {\bar R}_{\mu\nu} (x) = - (\partial_\mu \bar C_\nu - \partial_\nu\,\bar C_\mu)(x), \quad B^{(2)}_\mu (x) = - \partial_\mu \bar\beta (x), \quad
\bar f_1 (x) = -  \rho (x), \nonumber\\  && B^{(1)}_\mu (x) =  {\bar B}_\mu (x), \qquad \bar f_4 (x) = 0, \qquad f_3(x) = - \lambda (x).
\end{eqnarray}
Substituting these values into the super-expansion (18), we obtain 
\begin{eqnarray}
&& {\tilde B}^{(ab)}_{\mu\nu} (x, \theta) = B_{\mu\nu}(x) + \theta\, \Big[- (\partial_\mu \bar C_\nu - \partial_\nu\,\bar C_\mu)(x) \Big]
\equiv B_{\mu\nu}(x) + \theta\,\Big( s_{ab}\, B_{\mu\nu}(x) \Big),\nonumber\\
&& {\tilde {\cal F}}^{(ab)}_\mu (x, \theta) = C_\mu(x) + \theta\, \Big( {\bar B}_\mu  \Big) \equiv C_\mu(x) + \theta\,\Big( s_{ab}\, C_\mu (x) \Big),\nonumber\\
&& \tilde {\bar {\cal F}}^{(ab)}_\mu (x, \theta) = \bar C_\mu(x) + \theta\,\Big(  - \partial_\mu \bar\beta  \Big) \equiv 
\bar C_\mu(x) + \theta\,\Big( s_{ab}\, \bar C_\mu (x) \Big),\nonumber\\
&&  \tilde \Phi^{(ab)}_1 (x, \theta) = \phi_1 (x) + \theta\, \Big( -  \rho (x) \Big) \equiv  \phi_1 (x) + \theta\, \Big( s_{ab} \phi_1 (x) \Big),\nonumber\\
&& \tilde \Phi^{(ab)}_2 (x, \theta) = \phi_2 (x) + \theta \, \Big( 0 \Big) \equiv \phi_2 (x) + \theta \, \Big( s_{ab}\, \phi_2 (x) \Big), \nonumber\\
&& \tilde \beta^{(ab)} (x, \theta) = \beta (x) + \theta \,\Big( - \lambda (x) \Big) \equiv \beta (x) + \theta \,\Big( s_{ab}\, \beta (x) \Big), \nonumber\\
&& \tilde {\bar\beta}^{(ab)} (x, \theta) = \bar\beta (x) + \theta\, \Big( 0 \Big) \equiv \bar\beta (x) + \theta\, \Big(s_{ab}\, \bar\beta (x)\Big),
\end{eqnarray}
where the superscript ($ab$), on the chiral superfields, denotes the expansion on the superfields after the application of the anti-BRST 
invariant restrictions [cf. (30)] on the (super)fields of our present theory.

We offer some comments on the relationships that have been obtained in Eqn. (31). It is straightforward to note that
the first row of the restrictions in (30) produces the results: $\bar f_4 (x) = 0, \, \rho (x) \bar f_1 (x) = 0, \, 
\lambda(x) \bar f_3 (x) = 0$. These relatioships imply that the non-trivial solutions are $\bar f_1 (x) \propto \rho (x)$
and $\bar f_3 (x) \propto \lambda (x)$. However, the second and third row restrictions in (30) are satisfied if and only 
if we take into account: $B^{(1)}_\mu (x) = {\bar B}_\mu (x), \, \bar f_1 (x) = - \rho (x),\, 
B^{(2)}_\mu (x) = - \partial_\mu \bar\beta $ and $\bar f_3 (x) = - \lambda (x)$. Finally, when we focus on the {\it last}
anti-BRST invariant restriction in (30), we have to take into account all the  inputs that have been derived from 
the restrictions in the first, second and third rows of (30). Substitutions of all these inputs, ultimately,
leads to the derivation of $\bar R_{\mu\nu}(x) = - (\partial_\mu \bar C_\nu - \partial_\nu \bar C_\mu)$ in the expansion of
${\tilde{\cal B}}_{\mu\nu} (x, \theta)$.

We end this section with the remark that the geometrical interpretation of the  invariance [cf. (8), (11)] of the Lagrangian densities
${\cal L}_{(B, {\cal B})}$ and ${\cal L}_{(\bar B, {\bar{\cal B}})}$  can also be captured within the framework
of augmented (anti-)chiral superfield formalism. It is straightforward to note that these 
ordinary Lagrangian densities can be generalized onto the (anti-)chiral
super-submanifolds as their counterpart (anti-)chiral super Lagrangian densities, namely;
\begin{eqnarray}
{\cal L}_{(\bar B, {\bar{\cal B}})} \longrightarrow {\tilde {\cal L}}^{(ad)}_{(\bar B, {\bar{\cal B}})} 
&=& \frac{1}{2}\bar{\cal B}\cdot \bar{\cal{B}} - \bar{\cal B}^{\mu}\Bigl(\frac{1}{2}\varepsilon_{\mu\nu\eta\kappa}\partial^{\nu} {\tilde B}^{\eta\kappa(ad)} 
- \partial_{\mu} {\tilde \Phi}^{(ad)}_{2} \Bigr) \nonumber\\
&+&  {\bar B}^{\mu} \Bigl(\partial^{\nu} {\tilde B}^{(ad)}_{\nu\mu} 
- \partial_{\mu} {\tilde \Phi}^{(ad)}_{1} \Bigr)-\frac{1}{2}\bar B\cdot \bar{B} +  \partial_{\mu} \tilde{\bar{\beta}}^{(ad)}\partial^{\mu} 
{\tilde \beta}^{(ad)}  \nonumber\\
&+&  \Bigl( \partial_{\mu} \tilde{\bar {\cal F}}^{(ad)}_{\nu} - \partial_{\nu}\tilde{\bar {\cal F}}^{(ad)}_{\mu} \Bigr)\,(\partial^{\mu} 
\tilde {\cal F}^{\nu (ad)}) \nonumber\\ 
 &+& \Bigl(\partial \cdot {\tilde{\cal F}}^{(ad)} - \lambda \Bigr)\,\rho + \Bigl(\partial \cdot \tilde{{\bar {\cal F}}}^{(ad)} 
+ \rho \Bigr)\,\lambda, \nonumber\\
{\cal L}_{(\bar B, {\bar{\cal B}})} \longrightarrow {\tilde {\cal L}}^{(ab)}_{(\bar B, {\bar{\cal B}})} &=& \frac{1}{2}\bar{\cal B}\cdot \bar{\cal{B}} 
- \bar{\cal B}^{\mu}\Bigl(\frac{1}{2}\varepsilon_{\mu\nu\eta\kappa}\partial^{\nu} {\tilde B}^{\eta\kappa(ab)} 
- \partial_{\mu}{\tilde \Phi}^{(ab)}_{2} \Bigr) \nonumber\\
 &+&  {\bar B}^{\mu} \Bigl(\partial^{\nu} {\tilde B}^{(ab)}_{\nu\mu} - \partial_{\mu} {\tilde \Phi}^{(ab)}_{1} \Bigr)  
- \frac{1}{2}\bar B\cdot \bar{B} +  \partial_{\mu} \tilde{\bar{\beta}}^{(ab)}\partial^{\mu} {\tilde \beta}^{(ab)} \nonumber\\ 
&+&  \Bigl(\partial_{\mu} \tilde{\bar {\cal F}}^{(ab)}_{\nu} 
- \partial_{\nu}\tilde{\bar {\cal F}}^{(ab)}_{\mu} \Bigr)\,(\partial^{\mu} \tilde {\cal F}^{\nu (ab)}) \nonumber\\
 &+&  \Bigl(\partial \cdot {\tilde{\cal F}}^{(ab)} - \lambda \Bigr)\,\rho + \Bigl(\partial \cdot \tilde{{\bar {\cal F}}}^{(ab)} 
+ \rho \Bigr)\,\lambda, \nonumber\\
{\cal L}_{(B, {\cal B})} \longrightarrow {\tilde {\cal L}}^{(d)}_{( B, {\cal B})} &=& \frac{1}{2} {\cal B}\cdot {\cal{B}}
 - {\cal B}^{\mu}\Bigl(\frac{1}{2}\varepsilon_{\mu\nu\eta\kappa}\partial^{\nu} {\tilde B}^{\eta\kappa(d)} 
+ \partial_{\mu} {\tilde \Phi}^{(d)}_{2} \Bigr) \nonumber\\
 &+&   B^{\mu}\Bigl(\partial^{\nu} {\tilde B}^{(d)}_{\nu\mu} + \partial_{\mu} {\tilde \Phi}^{(d)}_{1} \Bigr) 
- \frac{1}{2}\bar B\cdot \bar{B} +  \partial_{\mu} \tilde{\bar{\beta}}^{(d)}\partial^{\mu} {\tilde \beta}^{(d)}  \nonumber\\
&+&  \Bigl(\partial_{\mu} \tilde{\bar {\cal F}}^{(d)}_{\nu} 
- \partial_{\nu}\tilde{\bar {\cal F}}^{(d)}_{\mu}\Bigr)\,(\partial^{\mu} \tilde {\cal F}^{\nu (d)}) \nonumber\\ 
 &+&  \Bigl(\partial \cdot {\tilde{\cal F}}^{(d)} - \lambda\Bigr)\,\rho + \Bigl(\partial \cdot \tilde{{\bar {\cal F}}}^{(d)} + \rho \Bigr)\,\lambda, \nonumber\\
{\cal L}_{(B, {\cal B})} \longrightarrow {\tilde {\cal L}}^{(b)}_{(( B, {\cal B})}&=& \frac{1}{2}{\cal B}\cdot {\cal{B}}
 - {\cal B}^{\mu}\Bigl(\frac{1}{2}\varepsilon_{\mu\nu\eta\kappa}\partial^{\nu} {\tilde B}^{\eta\kappa(b)} + \partial_{\mu} {\tilde \Phi}^{(b)}_{2} \Bigr) \nonumber\\
&+&  B^{\mu}\Bigl(\partial^{\nu} {\tilde B}^{(b)}_{\nu\mu} + \partial_{\mu}{\tilde \Phi}^{(b)}_{1} \Bigr)  
- \frac{1}{2}\bar B\cdot \bar{B} +  \partial_{\mu} \tilde{\bar{\beta}}^{(b)}\partial^{\mu} {\tilde \beta}^{(b)}  \nonumber\\
&+&  \Bigl(\partial_{\mu} \tilde{\bar {\cal F}}^{(b)}_{\nu} - \partial_{\nu}\tilde{\bar {\cal F}}^{(b)}_{\mu}\Bigr)\,(\partial^{\mu} \tilde {\cal F}^{\nu (b)}) \nonumber\\ 
 &+&  \Bigl(\partial \cdot {\tilde{\cal F}}^{(b)} - \lambda\Bigr)\,\rho + \Bigl(\partial \cdot \tilde{{\bar {\cal F}}}^{(b)} + \rho \Bigr)\,\lambda, 
\end{eqnarray}
where the superscripts ($ad,\, ab,\, d,\, b$) denote that the superfield expansions (22), (32), (17) and (28) have been taken into account.
Now, it is elementary to check that we have the following explicit restrictions
\begin{eqnarray}
\frac{\partial}{\partial \theta}{\tilde {\cal L}}^{(ad)}_{(\bar B, {\bar{\cal B}})} = \frac{\partial}{\partial x^\mu} \Big[ (\partial^{\mu}C^{\nu} 
- \partial^{\nu}C^{\mu})\, {\bar{\cal B}}_\nu + \rho\,\partial^{\mu}\beta + \lambda \,{\bar{\cal B}}^\mu  \Big]
\equiv s_{ad}\,{\cal L}_{(\bar B, \bar {\cal B})},\nonumber\\
\frac{\partial}{\partial \theta}{\tilde {\cal L}}^{(ab)}_{(\bar B, {\bar{\cal B}})} = \frac{\partial}{\partial x^\mu} \Big[ (\partial^{\mu}\bar{C^{\nu}} 
- \partial^{\nu}\bar{C^{\mu}})\bar{B_{\nu}} - \rho\, \bar{B^{\mu}} + \lambda \,\partial^{\mu}\bar{\beta}\Big]
\equiv s_{ab}\,{\cal L}_{(\bar B, \bar{\cal B})},\nonumber\\
\frac{\partial}{\partial \bar\theta} {\tilde {\cal L}}^{(d)}_{( B, {\cal B})} =  \frac{\partial}{\partial x^\mu} \Big[ (\partial^{\mu}\bar C^{\nu} 
- \partial^{\nu}{\bar C}^\mu){\cal B}_{\nu} - \lambda\,\partial^{\mu} {\bar \beta} - \rho\, {\cal B}^{\mu} \Big]
\equiv  s_d\,{\cal L}_{(B, {\cal B})}, \nonumber\\
\frac{\partial}{\partial \bar\theta} {\tilde {\cal L}}^{(b)}_{( B, {\cal B})} =  \frac{\partial}{\partial x^\mu} \Big[(\partial^{\mu}C^{\nu} - \partial^{\nu}C^{\mu})B_{\nu} + \rho\, \partial^{\mu}\beta + \lambda\, B^{\mu} \Big] \equiv s_b\,{\cal L}_{(B, {\cal B})},
\end{eqnarray}
which provide the geometrical interpretation for the invariance(s) of the Lagrangian densities  (1) and (2) within the framework of our
augmented (anti-) chiral superfield formalism. It states that the translation of the {\it sum} of a specific
combination of composite (anti-)chiral (super)fields, 
present in the super Lagrangian densities (33) along the $(\theta)\bar\theta$-directions of the chiral and anti-chiral 
super-submanifolds, produces the ordinary spacetime derivatives (8), (10), (9) and (11). This observation, it turn, implies the 
(anti-)co-BRST and (anti-)BRST invariance of the action integrals corresponding to the Lagrangian densities (1) and (2) for the 
physical fields that vanish off at infinity.

\section{Conserved Fermionic Charges: Nilpotency and Absolute Anticommutativity Properties}
\noindent
According to the celebrated Noether theorem, it is evident that the nilpotent (anti-)BRST and (anti-)co-BRST
symmetry transformations (which are infinitesimal and continuous) lead to the derivation of Noether conserved currents and charges.
These currents and charges have been derived in our earlier work [23]. We list here the explicit
mathematical forms of the above conserved fermionic charges as:
\begin{eqnarray}
Q_{ab}& = &\,\int\,d^3x\,\Bigl[\rho\,\bar B^0 - (\partial^0\bar C^i - \partial^i\bar C^0)\,\bar B_i 
- \,(\partial^0 C^i - \,\partial^i C^0)\,\partial_i\,\bar\beta \nonumber \\
& - &\,\lambda\,\partial^0\,\bar\beta 
- \,\varepsilon^{ijk}(\partial_i \bar C_j)\,\bar {\cal B}_k \Bigr], \nonumber \\
Q_{b}& = &\,\int\,d^3x\,\Bigl[(\partial^0\bar C^i - \partial^i\bar C^0)\,\partial_i\beta 
- \,(\partial^0 C^i - \,\partial^i C^0)\,B_i -\,\lambda\,B^0 \nonumber \\
& - &\,\rho\,\partial^0\beta
- \,\varepsilon^{ijk}(\partial_i C_j)\,{\cal B}_k \Bigr], \nonumber \\
Q_{ad}& = &\,\int\,d^3x\,\Bigl[(\partial^0 C^i - \partial^i C^0)\,\bar {\cal B}_i - \,(\partial^0\bar C^i 
- \partial^i\bar C^0)\,\partial_i \beta  \nonumber \\
& + &\,\lambda\,\bar{\cal B}^0 + \,\rho\,\partial^0\beta
- \,\varepsilon^{ijk}(\partial_i  C_j)\,\bar  B_k \Bigr], \nonumber \\
Q_{d}& = &\,\int\,d^3x\,\Bigl[(\partial^0 \bar C^i - \partial^i\bar C^0)\,{\cal B}_i - \,(\partial^0 C^i 
- \partial^i C^0)\,\partial_i \bar\beta  \nonumber \\
& - &\,\rho\,{\cal B}^0 -\,\lambda\,\partial^0\bar\beta - \,\varepsilon^{ijk}(\partial_i  \bar C_j)\, B_k \Bigr].
\end{eqnarray}
Our central aim, in this section, is to prove the nilpotency and absolute anticommutativity properties of the above charges. Thus, we express
these charges in the forms where, first of all, their nilpotency property becomes (obvious in a straightforward fashion). In
this respect, we use the following Euler-Lagrange (EL) equations of motion (EOM)
\begin{eqnarray}
&&\partial_\mu\,(\partial^\mu\,C^\nu - \partial^\nu\,C^\mu) = -\,\partial^\nu \,\lambda \quad \Longrightarrow  \quad
\partial_i\,(\partial^i\, C^0 - \partial^0\, C^i) = - \dot\lambda, \nonumber \\
&&\partial_\mu\,(\partial^\mu\,\bar C^\nu - \partial^\nu\,\bar C^\mu) = \partial^\nu \,\rho \qquad \Longrightarrow \quad
\partial_i\,(\partial^i\, \bar C^0 - \partial^0\, \bar C^i) = \dot\rho,
\end{eqnarray}
which are derived from the Lagrangian densities (1) and /or (2). Thus, the convenient forms of the charges (35)
become:
\begin{eqnarray}
Q_{ab} &=& \int\,d^3 x \,\Bigl[\rho\,\bar B^0 - (\partial^0 \bar C^i - \partial^i \bar C^0)\, \bar B_i
+ {\dot \lambda} \bar \beta \nonumber\\ 
&-& \lambda\,\dot{\bar \beta} - \varepsilon^{ijk}(\partial_i \bar C_j)\,\bar {\cal B}_k \Bigr],\nonumber\\
Q_{b} &=& \int\,d^3 x \,\Bigl[{\dot \rho}\,\beta - (\partial^0 C^i - \partial^i C^0)\,B_i - \lambda\, B^0 \nonumber\\
&-& \rho\,{\dot\beta} - \varepsilon^{ijk}(\partial_i C_j)\,{\cal B}_k\Bigr], \nonumber\\
Q_{ad} &=& \int\,d^3 x \Bigl[(\partial^0 C^i - \partial^i\, C^0)\, \bar {\cal B}_i - \dot\rho\,\beta \nonumber\\
&+& \lambda\,\bar {\cal B}^0 + \rho\,\dot\beta - \varepsilon^{ijk}(\partial_i C_j)\,\bar B_k \Bigr], \nonumber \\
Q_d &=& \int\,d^3 x\,\Bigl[(\partial^0\,\bar C^i - \partial^i\,\bar C^0)\,{\cal B}_i
+ \dot\lambda\,\bar\beta - \rho\,{\cal B}^0 \nonumber\\
&-& \lambda\, \dot{\bar\beta}- \varepsilon^{ijk}(\partial_i \bar C_j)\, B_k \Bigr].
\end{eqnarray}
The above expressions of the conserved charges can be written in the following (anti-)BRST and (anti-)co-BRST
{\it exact} forms\footnote{In writing of these {\it exact} forms, our knowledge of superfield formalism has been very much
helpful. Thus, we would like to lay emphasis on the fact that the content and essence of Eqns. (38), (40) and (41) are {\it dependent} on one-another.
In other words, the knowledge of the {\it ordinary} spacetime symmetry transformations in the ordinary space and their structure (cf. Eqn. (38)) helps us in deriving the appropriate equations in the {\it superspace} with their proper forms (cf. Eqns. (40) and (41)) and {\it vi\'ce-versa}.}:
\begin{eqnarray}
Q_{ab}& = &s_{ab}\, \int d^3 x\,\Bigl[B^{0i}\,\bar B_i - \bar \beta\,\dot \beta - \phi_1\,\bar B^0
+\dot{\bar\beta}\,\beta + \frac{1}{2}\,\varepsilon^{ijk}\, B_{ij}\,\bar {\cal B}_k\Bigr], \nonumber \\
Q_{b}& = &s_b\, \int d^3 x\,\Bigl[B^{0i} B_i - \dot {\bar \beta}\,\beta + \phi_1\, B^0 + \bar \beta\,{\dot \beta}
+ \frac {1}{2}\,\varepsilon^{ijk}\, B_{ij}\,{\cal B}_k \Bigr], \nonumber \\
Q_{ad}& = &s_{ad}\, \int d^3 x \, \Bigl[\frac{1}{2}\,\varepsilon^{ijk}\,\bar {\cal B}_i\,B_{jk} -\dot {\bar \beta}\,\beta
+ \phi_2\,\bar {\cal B}^0 - \,\rho\, C^0 +\, B^{0i}\,\bar B_i \Bigr], \nonumber \\
Q_{d} & = &s_d\, \int d^3 x\,\Bigl[\frac{1}{2}\,\varepsilon^{ijk}\, {\cal B}_i \,B_{jk} - \bar \beta\,{\dot \beta} 
- \phi_2\,{\cal B}^0 - \lambda\,\bar C^0 + B^{0i}\, B_i\Bigr].
\end{eqnarray}
It is now straightforward to note that the nilpotency of the (anti-)BRST and (anti-)co-BRST
symmetries (i.e. $s_{(a)b}^2 = 0, \,s_{(a)d} = 0$) {\it imply} the following:
\begin{eqnarray}
&&s_{ab}^2 = 0 \quad\, \Longleftrightarrow \quad \;\, s_{ab}\, Q_{ab} = -\,i\,\{Q_{ab}, Q_{ab}\} = 0 
\quad \;\;\Longrightarrow \quad Q_{ab}^2 = 0,\nonumber \\
&&s_{b}^2 = 0 \quad\;\,\Longleftrightarrow \quad \;\, s_{b}\, Q_{b} = -\,i\,\{Q_{b}, Q_{b}\} = 0 
\qquad  \quad \, \Longrightarrow \quad  Q_b^2 = 0,\nonumber \\ 
&&s_{ad}^2 = 0 \quad\,\Longleftrightarrow \quad \;\, s_{ad}\, Q_{ad} = -\,i\,\{Q_{ad}, Q_{ad}\} = 0 
\quad \,\Longrightarrow \quad  Q_{ad}^2 = 0, \nonumber   \\ 
&&s_{d}^2 = 0 \quad\;\;\Longleftrightarrow \quad \;\, s_{d}\, Q_{d} = -\,i\,\{Q_{d}, Q_{d}\} = 0 \qquad \;\;\Longrightarrow 
\quad  Q_d^2 = 0,   
\end{eqnarray}
where we have used the basic concepts behind the continuous symmetries and their generators
(which are nothing but the conserved charges).
In other words, we observe that the nilpotency ($s_{(a)b}^2 = 0,\; s_{(a)d}^2 = 0$) of the (anti-) BRST
and (anti-)co-BRST symmetries are very intimately connected with the nilpotency (i.e. $Q_{(a)b}^2 = 0, \; Q_{(a)d}^2 = 0$)
of the corresponding conserved (anti-)BRST ($Q_{(a)b}$) and (anti-)co-BRST charges ($Q_{(a)d}$).

We now concentrate on expressing the above observations in the language of the translational
generators along the Grassmannian directions of the (anti-)chiral super-submanifolds of the
general (4, 2)-dimensional supermanifold and the (anti-)chiral superfields that have been obtained after the application of BRSTIRs and
CBRSTIRs. In this context, we have to utilize the super expansions (32) and (28) to express the (anti-)BRST charges $Q_{(a)b}$
in terms of the (anti-)chiral superfields and Grassmannian differentials and/or derivatives. For instance, we 
observe that the (anti-)BRST charges ($Q_{(a)b}$) (cf. Eqn. (38)) can be written, in their explicit form, as:
\begin{eqnarray}
Q_{ab}& = &\frac{\partial}{\partial\theta}\,\int d^3 x\,\Big[\tilde {B}^{0i(ab)}(x, \theta)\,\bar B_i(x) 
- \dot {\tilde\beta}^{(ab)}(x, \theta)\,\bar \beta(x) - \tilde {\Phi}_1^{(ab)}(x, \theta)\,\bar B^0(x)\nonumber \\
& + &\tilde {\beta}^{(ab)} (x, \theta)\,\dot{\bar\beta}(x) 
+ \frac{1}{2}\,\varepsilon^{ijk}\,\tilde  {B}_{ij}^{(ab)}(x, \theta)\,{\bar{\cal B}}_k(x)\Big] \nonumber \\
& \equiv &\int d\theta\,\int d^3 x\,\Bigl[\tilde {B}^{0i (ab)}(x, \theta)\,\bar B_i(x) 
+ \dot {\tilde\beta}^{(ab)}(x, \theta)\,\bar \beta(x) - \tilde {\Phi}_1^{(ab)}(x, \theta)\,\bar B^0(x)\nonumber \\
& + &\tilde {\beta}^{(ab)} (x, \theta)\,\dot{\bar\beta}(x) 
+ \frac{1}{2}\,\varepsilon^{ijk}\,\tilde  {B}_{ij}^{(ab)}(x, \theta)\,{\bar{\cal B}}_k(x)\Bigr], \nonumber \\
Q_{b} & = &\frac{\partial}{\partial\bar\theta}\,\int d^3 x\,\Big[{\tilde B}^{0i(b)}(x, \bar\theta)\, B_i(x) 
- \dot {\tilde{\bar\beta}}^{(b)}(x, \bar\theta)\, \beta(x) + \tilde {\Phi}_1^{(b)}(x, \bar\theta)\, B^0(x)\nonumber \\
& + &\tilde {\bar\beta}^{(b)} (x, \bar\theta)\,\dot{\beta}(x) 
+ \frac{1}{2}\,\varepsilon^{ijk}\,\tilde  {B}_{ij}^{(b)}(x, \bar\theta)\,{\cal B}_k(x)\Big] \nonumber \\
& \equiv &\int d\bar\theta\,\int d^3 x\,\Big[\tilde {B}^{0i(b)}(x, \bar\theta)\, B_i(x) 
- \dot {\tilde{\bar\beta}}^{(b)}(x, \bar\theta)\, \beta(x) + \tilde {\Phi}_1^{(b)}(x, \bar\theta)\, B^0(x)\nonumber \\
& + &\tilde {\bar\beta}^{(b)} (x, \bar\theta)\,\dot{\beta}(x) 
+ \frac{1}{2}\,\varepsilon^{ijk}\,\tilde  {B}_{ij}^{(b)}(x, \bar\theta)\,{\cal B}_k(x)\Big]. 
\end{eqnarray}
Thus, we note that we have to  use the ordinary fields (basic as well as auxiliary) and the (anti-)chiral superfields
(obtained after BRSTIRs) to express the (anti-)BRST charges in the language of (anti-)chiral superfield approach to BRST
formalism where the Grassmannian derivatives/differential are also exploited in a judicious manner. It is quite
clear to observe that the nilpotency of ($\partial_\theta,\, \partial_{\bar\theta}$) is intimately connected with
the nilpotency of the (anti-) BRST charges because we already see that 
$\partial_\theta\,Q_{ab} = 0,\,\partial_{\bar\theta}\,Q_b = 0$.

We can express the (anti-)co-BRST charges in exactly similar fashion as we have done for the
(anti-)BRST charges. Towards this goal in mind,  we observe that the following are true, namely;
\begin{eqnarray}
Q_{ad}& = &\frac{\partial}{\partial\theta}\int d^3 x\,\Bigl[ \frac{1}{2}\varepsilon^{ijk}\bar{\cal B}_i(x)\,
{\tilde B}_{jk}^{(ad)}(x, \theta) - \dot{\tilde{\bar\beta}}^{(ad)}(x, \theta)\,\beta(x)\nonumber \\
&+& {\tilde\Phi}_2^{(ad)}(x, \theta)\,\bar{\cal B}^0(x)
 -  \rho(x)\,{\tilde {\cal F}}^{(0)(ad)}(x, \theta) + \tilde B^{0i(ad)}(x, \theta)\,B_i(x)\Bigr], \nonumber \\
&\equiv &\int d\theta\,\int d^3x\,\Bigl[ \frac{1}{2}\varepsilon^{ijk}\bar{\cal B}_i(x)\,
{\tilde B}_{jk}^{(ad)}(x, \theta) - \dot{\tilde{\bar\beta}}^{(ad)}(x, \theta)\,\beta(x)\nonumber \\
&+& {\tilde\Phi}_2^{(ad)}(x, \theta)\,\bar{\cal B}^0(x)
 -  \rho(x)\,{\tilde {\cal F}}^{(0)(ad)}(x, \theta) + \tilde B^{0i(d)}(x, \theta)\,B_i(x)\Bigr], \nonumber \\
Q_{d}& = &\frac{\partial}{\partial\bar\theta}\int d^3 x\,\Bigl[ \frac{1}{2}\varepsilon^{ijk}{\cal B}_i(x)\,
{\tilde B}_{jk}^{(d)}(x, \bar\theta) - \dot{\tilde{\beta}}^{(d)}(x, \bar\theta)\,\bar\beta(x) \nonumber \\
&-& {\tilde\Phi}_2^{(d)}(x, \bar\theta)\,{\cal B}^{0}(x)
- \lambda(x)\, \tilde {\bar{\cal F}}^{0(d)}(x, \bar\theta) + {\tilde B}^{0i(d)}(x, \bar\theta)\,B_i(x)\Bigr], \nonumber \\
&\equiv &\int d\bar\theta\,\int d^3 x\,\Bigl[ \frac{1}{2}\varepsilon^{ijk}{\cal B}_i(x)\,
{\tilde B}_{jk}^{(d)}(x, \bar\theta) - \dot{\tilde{\beta}}^{(d)}(x, \bar\theta)\,\bar\beta(x)\nonumber \\
&-& {\tilde\Phi}_2^{(d)}(x, \bar\theta)\,{\cal B}^{0}(x)
- \lambda(x)\,\tilde {\bar{\cal F}}^{0(d)}(x, \bar\theta) + {\tilde B}^{0i(d)}(x, \bar\theta)\,B_i(x)\Bigr].
\end{eqnarray}
In the above expressions, we have utilized the superfield expansions (12) and (18) that have been
obtained after the application of CBRSTIRs. Further, we also note that the suitable ordinary fields of the 
Lagrangian densities (1) and/or (2) have {\it also} been used in expressing the above forms of the nilpotent
(anti-)co-BRST conserved charges. It is clear, from the above expressions, that 
$\partial_{\bar\theta}\,Q_{ad} = 0$ and $\partial_{\bar\theta}\,Q_{d} = 0$
due to the nilpotency (i.e. $\partial_\theta^2 = \partial_{\bar\theta}^2 = 0 $) of the Grassmannian
translational generators $\partial_\theta$ and $\partial_{\bar\theta}$. In other words, we observe that 
the nilpotency of the (anti-)co-BRST symmetries (and their corresponding charges) is deeply connected with
the nilpotency ($\partial_\theta^2 = \partial_{\bar\theta}^2 = 0$) of the translational generators $\partial_\theta$ and $\partial_{\bar\theta}$. These observations, ultimately, are connected with the nilpotency ($Q^2_{(a)d} = 0$) of the (anti-) co-BRST charges.

We now concentrate on the proof of the absolute anticommutativity properties of the
off-shell nilpotent (anti-)BRST and (anti-)co-BRST charges within the framework of our present 
augmented (anti-)chiral superfield approach to BRST formalism. Towards this goal in mind,
we express the fermionic (anti-)BRST conserved charges as follows
\begin{eqnarray}
 Q_{ab} &=& \int d^3 x\,\Big[ \dot \lambda \bar \beta - \lambda \dot {\bar \beta} + \rho B^0 + \rho \dot {\phi_1} - \dot \rho \phi_1 
- (\partial^0 {\bar C}^i -  \partial^0 {\bar C}^i) B_i \nonumber\\ 
&+& (\partial^0  B^i - \partial^i  B^0) \bar C_i \Big ],\nonumber\\
 Q_b &=& \int d^3 x\,\Big[ \dot \rho \beta - \dot \lambda \phi_1 + \lambda \dot \phi_1 - \rho \dot\beta - \lambda \bar B^0
- (\partial^0 C^i - \partial^i C^0)\bar B_i \nonumber\\
&+& (\partial^0 \bar B^i - \partial^i \bar B^0) C_i \Big],
\end{eqnarray}
where we have used the equations of motion (36) and CF-condition (7) in addition to the EOMs:
$\varepsilon ^{ijk}\partial_j{\cal B}_k = -(\partial^0 B^i - \partial^i B^0)$ 
and $\varepsilon ^{ijk}\partial_j \bar{\cal B}_k = -(\partial^0 \bar B^i - \partial^i \bar B^0)$ etc. (see, e.g. [23] for details).
The above forms of the (anti-)BRST charges can be expressed, in terms of the
nilpotent (anti-)BRST transformations $s_{(a)b}$ (cf. Eqns. (5), (6)), as quoted below:
\begin{eqnarray}
&& Q_{ab} = s_b\,\int d^3 x \Big[\phi_1 \dot {\bar\beta }- \dot \phi_1 \bar\beta - \bar\beta B^0 - (\partial^0 {\bar C}^i 
- \partial^i {\bar C}^0) {\bar C}_i \Big],  \nonumber\\
&& Q_b = s_{ab}\,\int d^3 x \Big[ \phi_1 \dot \beta - \dot \phi_1 \beta + \beta {\bar B}^0 + (\partial^0 C^i - \partial^i C^0) C_i \Big].
\end{eqnarray}
The above expressions provide the proof of the absolute anticommutativity property of the
(anti-)BRST charges in the ordinary 4$D$ Minkowskian spacetime.
This statement can be corroborated in a mathematical language in the following fashion
\begin{eqnarray}
s_b Q_{ab} = -i\,\{ Q_{ab}, Q_b \} = 0 \quad \Longleftrightarrow  \quad s_b^2 = 0,  \nonumber \\
s_{ab} Q_b = -i\, \{Q_b, Q_{ab} \} = 0 \quad \Longleftrightarrow \quad s_{ab}^2 = 0,
\end{eqnarray}
where we have used the basic concepts behind the continuous symmetries and their generators (as the Noether
conserved charges corresponding to these very continuous symmetry transformations). From Eq. (44), it is 
very much clear that the absolute anticommutativity property of the conserved charges is very intimately
connected to the {\it nilpotency} of the continuous symmetry transformations they generate.

We are in the position now, to capture the above absolute anticommutating property of the conserved charges in the terminology
of the augmented (anti-)chiral superfield approach to BRST formalism. Towards the goal
in mind, we note the following:
\begin{eqnarray}
Q_{ab} &=& \frac{\partial}{\partial {\bar\theta}}\,\int d^3 x \Big[{\tilde \Phi}^{(b)}_1 (x, \bar\theta) \dot {\tilde {\bar\beta}}^{(b)}(x, \bar\theta)
- \dot{ {\tilde \Phi}}^{(b)}_1 (x, \bar\theta) {\tilde {\bar\beta }}^{(b)}(x, \bar\theta)\nonumber\\ 
&-& {\tilde {\bar\beta}}^{(b)} (x, \bar\theta) B^0 (x) 
- \Big(\partial^0 {\tilde{\bar {\cal F}}}^{i(b)}(x, \bar\theta) - \partial^i {\tilde{\bar {\cal F}}}^{0(b)}(x, \bar\theta) \Big) {\tilde{\bar {\cal F}}}^{(b)}_i  (x, \bar\theta)\Big],  \nonumber\\
&\equiv& \int d \bar\theta \,\int d^3 x \Big[{\tilde \Phi}^{(b)}_1 (x, \bar\theta) \dot {\tilde {\bar\beta}}^{(b)}(x, \bar\theta)
- \dot{ {\tilde \Phi}}^{(b)}_1 (x, \bar\theta) {\tilde {\bar\beta }}^{(b)}(x, \bar\theta) \nonumber\\
&-& {\tilde {\bar\beta}}^{(b)} (x, \bar\theta) B^0 (x) 
- \Big(\partial^0 {\tilde{\bar {\cal F}}}^{i(b)}(x, \bar\theta) - \partial^i {\tilde{\bar {\cal F}}}^{0(b)}(x, \bar\theta) \Big) {\tilde{\bar {\cal F}}}^{(b)}_i  (x, \bar\theta)\Big],  \nonumber\\
Q_b &=& \frac{\partial}{\partial \theta}\,\int d^3 x \Big[{ \tilde \Phi}^{(ab)}_1 (x, \theta) \dot {\tilde \beta}^{(ab)} (x, \theta) 
- \dot {\tilde \Phi}^{(ab)}_1 (x, \theta) \tilde\beta^{(ab)}(x, \theta) \nonumber\\ 
&+& \tilde \beta^{(ab)}(x, \theta) {\bar B}^0 (x)  
+ \Big(\partial^0 {\tilde{\cal F}}^{i(ab)} (x, \theta)- \partial^i {\tilde{\cal F}}^{0(ab)} (x, \theta) \Big) {\tilde{\cal F}^{(ab)}}_i (x, \theta) \Big] \nonumber\\
&\equiv& \int d \theta \,\frac{\partial}{\partial \theta}\,\int d^3 x \Big[{ \tilde \Phi}^{(ab)}_1 (x, \theta) \dot {\tilde \beta}^{(ab)} (x, \theta) 
- \dot {\tilde \Phi}^{(ab)}_1 (x, \theta) \tilde\beta^{(ab)}(x, \theta) \nonumber\\ 
&+& \tilde \beta^{(ab)}(x, \theta) {\bar B}^0 (x)  
+ \Big(\partial^0 {\tilde{\cal F}}^{i(ab)} (x, \theta)- \partial^i {\tilde{\cal F}}^{0(ab)} (x, \theta) \Big) {\tilde{\cal F}^{(ab)}}_i (x, \theta) \Big].\qquad
\end{eqnarray}
From the above expressions for $Q_{(a)b}$ (in the language of (anti-)chiral superfields obtained after appropriate
restrictions and Grassmannian directions and/or differential), it is evident that 
$\partial_\theta \, Q_b = 0,\; \partial_{\bar\theta} \, Q_{ab} = 0 $ due to the nilpotency property
($\partial_\theta^2 = 0,\; \partial_{\bar\theta}^2 = 0$) associated with the translational generators 
($\partial_\theta,\; \partial_{\bar\theta}$) along the Grassmannian directions ($\theta,\;\bar\theta$) of chiral
and anti-chiral super-submanifolds of the (4, 2)-dimensional supermanifold.

We now focus on the discussion of the absolute anticommutativity property, associated with the
conserved (anti-)co-BRST
charges, within the framework of augmented version of (anti-)chiral superfield formalism. 
In this connection, using the CF-type restrictions 
(${\cal B}_\mu - \bar{\cal B}_\mu = - \partial_\mu \phi_2$), the equations of motions (36) and other equations of motions
derived from the Lagrangian densities (1) and (2) (see, e.g. [23] for details), the (anti-)co-BRST
conserved charges ($Q_{(a)d}$) can be expressed as:
\begin{eqnarray}
 Q_{ad} &=& \int d^3 x\,\Big[\rho \dot \beta - \dot \rho \beta -  \dot \lambda \phi_2 + \lambda \dot {\phi_2} + \lambda {\cal B}^0 
 + (\partial^0 {\bar C}^i -  \partial^i {\bar C}^0) {\cal B}_i \nonumber\\ 
&-& (\partial^0  {\cal B}^i - \partial^i  {\cal B}^0) C_i \Big ],\nonumber\\
 Q_d &=& \int d^3 x\,\Big[ \dot \lambda \bar \beta -  \lambda \dot{\bar \beta} + \rho \dot \phi_2 - \dot \rho \phi_2
-\rho \bar {\cal B}^0 - (\partial^0 \bar C^i - \partial^i \bar C^0) \bar {\cal B}_i \nonumber\\
&+& (\partial^0 \bar{\cal B}^i - \partial^i \bar {\cal B}^0) \bar C_i \Big].
\end{eqnarray}
The above expressions for the conserved (anti-)co-BRST charges ($Q_{(a)d}$) can be written in the following co-exact forms
(using the (anti-)co-BRST symmetry transformations (3) and (4)), namely;
\begin{eqnarray}
&& Q_{ad} = s_d\,\int d^3 x \Big[\dot \beta \phi_2 - \beta \dot {\phi_2} - \beta {\cal B}^0 + (\partial^0  C^i - \partial^i  C^0) C_i \Big],  \nonumber\\
&& Q_d = s_{ad}\,\int d^3 x \Big[ \dot \phi_2 \bar \beta - \phi_1 \dot {\bar \beta} - \bar\beta \bar {\cal B}^0 
+ (\partial^0 \bar C^i - \partial^i \bar C^0) \bar C_i\Big].
\end{eqnarray}
From the above expressions, it is evident that the following observations are true by using the nilpotency 
($s^2_{(a)d} = 0$) property of ($s_{(a)d}$), namely;
\begin{eqnarray}
&& s_d Q_{ad} = -i\,\{ Q_{ad}, Q_d \} = 0 \qquad \; \Longleftrightarrow  \qquad s_d^2 = 0,  \nonumber \\
&& s_{ad} Q_d = -i\, \{ Q_d, Q_{ad} \} = 0 \qquad \; \Longleftrightarrow \qquad s_{ad}^2 = 0,
\end{eqnarray}
where we have used the basic concepts behind the continuous symmetries and their
generators. The latter are nothing but the conserved Noether charges that generate the continuous
symmetries. We lay emphasis on the fact that the nilpotency of the symmetry transformations are connected with the 
absolute anticommutativity of the conserved charges.

Within the framework of the augmented (anti-)chiral superfield approach, we can express the conserved
(anti-)co-BRST charges $Q_{(a)d}$ as follows:
\begin{eqnarray}
Q_{ad} &=& \frac{\partial}{\partial {\bar\theta}}\,\int d^3 x \Big[\dot{{\tilde \beta}}^{(d)} (x, \bar\theta)  {\tilde \Phi}_2^{(d)}(x, \bar\theta)
- \dot{ {\tilde \Phi}}^{(d)}_1 (x, \bar\theta) {\tilde {\bar\beta }}^{(d)}(x, \bar\theta)\nonumber\\
&-& {\tilde \beta}^{(d)} (x, \bar\theta) {\cal B}^0 (x) 
- \Big(\partial^0 {\tilde{\bar {\cal F}}}^{i(d)}(x, \bar\theta) - \partial^i {\tilde{\bar {\cal F}}}^{0(d)}(x, \bar\theta) \Big) {\tilde{\bar {\cal F}}}^{(d)}_i  (x, \bar\theta)\Big],  \nonumber\\
&\equiv & \int d \bar\theta \, \int d^3 x  \Big[\dot{{\tilde \beta}}^{(d)} (x, \bar\theta)  {\tilde \Phi}_2^{(d)}(x, \bar\theta)
- \dot{ {\tilde \Phi}}^{(d)}_1 (x, \bar\theta) {\tilde {\bar\beta }}^{(d)}(x, \bar\theta) \nonumber\\
&-& {\tilde \beta}^{(d)} (x, \bar\theta) {\cal B}^0 (x) 
- \Big(\partial^0 {\tilde{\bar {\cal F}}}^{i(d)}(x, \bar\theta) - \partial^i {\tilde{\bar {\cal F}}}^{0(d)}(x, \bar\theta) \Big) {\tilde{\bar {\cal F}}}^{(d)}_i  (x, \bar\theta)\Big],  \nonumber\\
Q_d &=& \frac{\partial}{\partial \theta}\,\int d^3 x \Big[{ \tilde \Phi}^{(ad)}_2 (x, \theta) \dot {\tilde {\bar \beta}}^{(ad)} (x, \theta) 
- \dot {\tilde \Phi}^{(ad)}_1 (x, \theta) {\tilde{\bar \beta}}^{(ad)}(x, \theta) \nonumber\\
&+& \tilde \beta^{(ad)}(x, \theta) {\bar {\cal B}}^0 (x)  
+ \Big(\partial^0 {\tilde{\cal F}}^{i(ad)} (x, \theta)- \partial^i {\tilde{\cal F}}^{0(ad)} (x, \theta) \Big) {\tilde{\cal F}^{(ad)}}_i (x, \theta) \Big] \nonumber\\
&\equiv& \int d \theta \int d^3 x  \Big[{ \tilde \Phi}^{(ad)}_2 (x, \theta) \dot {\tilde {\bar \beta}}^{(ad)} (x, \theta) 
- \dot {\tilde \Phi}^{(ad)}_1 (x, \theta) {\tilde{\bar \beta}}^{(ad)}(x, \theta) \nonumber\\
&+& {\tilde {\beta}}^{(ad)}(x, \theta) {\bar{\cal B}}^0 (x) 
+ \Big(\partial^0 {\tilde{\cal F}}^{i(ad)} (x, \theta)- \partial^i {\tilde{\cal F}}^{0(ad)} (x, \theta) \Big) {\tilde{\cal F}^{(ad)}}_i (x, \theta) \Big].\qquad
\end{eqnarray}
where we have used the (anti-)chiral superfields that have been obtained after the application
of appropriate restrictions as well as Grassmannian differentials and/or derivatives.
It is clear, from the above expressions for $Q_{(a)d}$ that $\partial_{\bar\theta} Q_{ad} = 0$ and 
 $\partial_{\theta} Q_{d} = 0$ due to the nilpotency property ($\partial_\theta^2 = 0, \partial_{\bar\theta}^2 = 0$)
 associated with the translational generators ($\partial_\theta, \partial_{\bar\theta}$) along the 
 Grassmannian directions ($\theta, \bar\theta $) of the (4, 1)-dimensional chiral
and anti-chiral super submanifolds of the (4, 2)-dimensional general supermanifold.

We end this section with the following remarks. First of all, it is evident from Eq. (23) that
$s_{(a)d}$ and $Q_{(a)d}$ are connected with the translational generators ($\partial_\theta, \partial_{\bar\theta}$) 
along the Grassmannian directions. This is why, the nilpotency ($\partial_\theta^2 = \partial_{\bar\theta}^2 = 0$)
and absolute anticommutativity ($\partial_\theta \partial_{\bar\theta} + \partial_{\bar\theta} \partial_\theta = 0 $)
of these generators are also intimately connected with the nilpotency
($Q_{(a)d}^2 = 0,\; Q_{(a)b}^2 = 0$) and absolute anticommutativity  
($Q_b\, Q_{ab} + Q_{ab}\, Q_b = 0,\; Q_d\, Q_{ad} + Q_{ad} \,Q_d = 0$) properties of the conserved fermionic charges
$Q_{(a)b}$ as well as $Q_{(a)d}$. A close look at (44) and (48) shows that the nilpotency property ($Q^2_{(a)b} = 0, \, Q^2_{(a)d} = 0$)
is a limiting case of absolute anticommutativity properties where $\{Q_b, Q_{ab}\} \equiv Q_b \, Q_{ab} + Q_{ab} \, Q_b =  0$ and
$\{Q_d, Q_{ad}\} \equiv Q_d \, Q_{ad} + Q_{ad} \, Q_d =  0$. This can be understood  more easily in the language of
translational generators along the Grassmannian directions where we have $\partial_\theta \partial_{\bar\theta} 
+ \partial_{\bar\theta} \partial_{\theta} \equiv \{\partial_\theta, \partial_{\bar\theta} \}  = 0$. If we set
$\partial_\theta = \partial_{\bar\theta}$ in the above relation for the absolute anticommutativity property, we obtain the 
nilpotency property ($\partial^2_\theta =  \partial^2_{\bar\theta} = 0$) of the translational generators 
($\partial_{\theta}, \partial_{\bar\theta}$) automatically.

\noindent
\section{Conclusions}

One of the key results of our present investigation is the {\it simplicity} of the theoretical technique that has been
used in the derivation of  (anti-)co-BRST and (anti-)BRST symmetry transformations for our present 
4$D$ {\it free} Abelian 2-form gauge theory which is a field theoretical model for the Hodge theory [23]. 
We note that we have not exploited the strength of the (dual-)HCs in our present derivation.
Rather, we have exploited the (anti-)co-BRST and (anti-)BRST invariance(s) to impose restrictions on the (anti-)chiral
superfields to achieve our goals in the sense that these conditions lead to the derivation of {\it proper}
(anti-) co-BRST and (anti-)BRST symmetry transformations. This should be contrasted with our earlier work [17]
where we have used the mathematical power and potential of the HC for the derivation of {\it only} nilpotent
(anti-)BRST symmetries for {\it this} 4$D$ field theoretic model of Hodge theory (i.e. 4$D$ free Abelian 2-form gauge theory).

It will be observed that we have utilized the (anti-)BRST invariance ($s_{(a)b}\, \phi_2 = 0$) 
of the field $\phi_2$ in (24). However, we have {\it not} touched upon the (anti-)BRST
invariance of the curvature tensor $H_{\mu\nu\eta}$ (i.e. $s_{(a)b}\, H_{\mu\nu\eta} = 0$) because this is connected
with the HC which has been {\it fully} utilized  in our earlier work [17]. One of the {\it novel} features of our  present investigation
is the observation that, even without the use of HC, one can derive the {\it proper} (anti-)BRST symmetries of our present theory
where {\it only} the property of (anti-)BRST invariance has been exploited extensively. In fact, the BRSTIRs have been motivated
by the physical arguments where we have demanded that the (anti-)BRST invariant quantities should be independent 
of the ``soul" coordinates $\theta$ and $\bar\theta$. In this connection, we note that 
the bosonic coordinates ($x^\mu$) of the superspace variable $Z^M = (x^\mu,\, \theta,\, \bar\theta)$ have been called as the ``body" 
coordinates and the pair of Grassmannian variables ($\theta,\, \bar\theta$) have been christened as the ``soul" coordinates in the older literature
(see, e.g. [24]). The former could be realized {\it physically} but the latter are only {\it mathematical} artifacts.
Thus, a physical quantity should be independent of the ``soul" coordinates. This is the physical input that 
has been incorporated in (C)BRSTIRs that have been invoked in the main body of our present text (because, in the discussion of 
a  model of the Hodge theory\footnote{A field theoretical model for the Hodge theory is the one whose symmetries  (and conserved charges)
provide the physical realizations of the de Rham cohomological operators of differential geometry. The 4$D$ free Abelian 2-form gauge theory
is one such  example within the framework of BRST formalism [23].},
the (anti-)co-BRST and (anti-)BRST invariant quantities are physical quantities).

The derivation of the proper (i.e. off-shell nilpotent and absolutely anticommuting) (anti-)co-BRST symmetries
is a {\it novel} result in our present investigation where we have not used the dual-HC. We have {\it not} utilized 
this idea of CBRSTIRs in our earlier work [17]  on the 4$D$ free Abelian 2-form gauge theory where we have used the superfield formalism
(with the HC) to derive {\it only} the (anti-)BRST symmetries. The highlight
of our present endeavor is the observation that the geometrical interpretations for the (anti-)BRST and (anti-)co-BRST symmetry transformations,
in terms of the translational generators $(\partial_\theta)\partial_{\bar\theta}$, remain the {\it same} 
as in our earlier works [9-13, 17]. As a consequence, the nilpotency property of the symmetry transformations {\it and} 
translational generators (along the Grassmannian directions) remain beautifully entangled and intertwined in an elegant manner 
(i.e. $s_{(a)b}^2 = 0, \,\, s_{(a)d}^2 = 0, \,\,\partial_{\theta}^2 = 0,\,\, \partial_{\bar\theta}^2 = 0)$ in our present endeavor, too.

A paragraph here about $s_{(a)d}\, (\partial ^\nu B_{\nu\mu}) = 0$.  
In our present endeavor, we have exploited the (anti-)co-BRST invariance of $\phi_1$
(i.e. $s_{(a)d}\, \phi_1 = 0$) in our physically motivated restriction (13).
However, we have {\it not} utilized the beauty and strength of the (anti-)co-BRST invariance of the 
gauge-fixing term [i.e. $s_{(a)d}\,(\partial ^\nu B_{\nu\mu}) = 0 $] which has its origin in the co-exterior 
derivative $\delta = -\, *\,d\,*$ where $*$ is the Hodge duality operation on the 4$D$ spacetime manifold [18-21]. To use 
this observation in the physical context, we have to develop the working-rule for the application of dual-HC in the context
of 4$D$ Abelian 2-form gauge theory as has been done in our earlier work on Abelian 1-form theory in 2$D$ and 4$D$ (see, e.g. [25] for details). 
We plan to pursue this direction of research in our future endeavor so that dual-HC could be
defined for our present 4$D$ Abelian 2-form gauge theory, too.

The importance of (anti-)BRST symmetries and corresponding charges is well-known. Here
we would like to dwell a bit on the physical importance of the (anti-)co-BRST symmetry transformations 
and corresponding conserved charges in the context of 1-form and 2-form gauge theories. In this context,
it is interesting to point out that, in our earlier work [26], we have established that the 2$D$ 
(non-)Abelian 1-form gauge theories (without any interaction with mater fields) belong to a new class of 
topological field theory (TFT) that captures a few aspects of Witten-type TFT [27] and some salient 
features of Shwarz-type TFT [28]. Furthermore, in our set of works on the free 4$D$ Abelian 2-form gauge theory
[29-32], we have exploited the (anti-)co-BRST symmetries and corresponding (anti-)co-BRST charges to
establish that  (i) this theory is a field theoretic model of Hodge theory where the Hodge decomposition
theorem could be applied (see, e.g. [23,29,30]), (ii) this theory is a model of quasi-TFT where the correct
recursion relations for the topological invariants exist [31], and (iii) the cohomological aspects 
of this theory can be discussed in the quantum Hilbert space with the (anti-)BRST, (anti-)co-BRST charges
and a unique bosonic charge (i.e. an appropriate anticommutator of the (anti-)BRST and (anti-)co-BRST charges) 
by choosing the {\it harmonic} state as the {\it physical} state that is annihilated by the (anti-)BRST
and (anti-)co-BRST charges {\it together} [32]. Thus, the (anti-)co-BRST symmetries and corresponding
conserved and nilpotent charges are physically very important in the context of discussion of  some key aspects
of {\it quantum} gauge theories within the
framework of BRST formalism (see, e.g. [29-32] for details).

The physically motivated restrictions on the (anti-)chiral superfields (that have been adopted in our present
endeavor) are very {\it general} because these can be used to derive any kind of {\it nilpotent} symmetry
transformations. In fact, we have exploited this idea (in an elegant manner) to derive the fermionic 
(i.e. nilpotent) symmetries for the ${\cal N} = 2$ supersymmetric (SUSY) quantum mechanical models in a
set of papers (see, e.g. [33-36]) and established that these models are physical examples of Hodge theory.
To be more precise, we have utilized the supersymmetric invariant restrictions on the
(anti-)chiral supervariables to derive the ${\cal N} = 2$ SUSY symmetries which are {\it nilpotent} (but
not absolutely anticommuting in nature). Thus, the observations made in our present endeavor are
quite simple and general which lead to the derivation of {\it nilpotent} symmetries for any kind of theory
(i.e. supersymmetric symmetries as well as (anti-)BRST and (anti-)co-BRST symmetries).

It would be a nice future problem to apply our
present idea to 2$D$ (non-) Abelian 1-form and 6$D$ Abelian 3-form gauge theories where the existence of (anti-)BRST 
and (anti-)co-BRST  transformations have been shown (in order to prove that these models are 
also the field theoretic examples of the Hodge theory).
We have {\it also} proven the 1$D$ system of rigid rotor [37] and 2$D$ self-dual fields [38] to be the examples of Hodge theory.
We plan to apply our present idea to these systems, too, so that our present technique could be put on a solid foundation.
In this context, it is gratifying to mention that we have already applied this idea to 1$D$ rigid rotor and 2$D$ self-dual
bosonic field theory and obtained the expected results [39,40]. In a very recent set of papers [41,42], we have exploited
our present ideas and shown the validity of absolute anticommutativity of conserved charges for the non-Abelian 1-form gauge 
theory (without any interaction with matter fields) and interacting Abelian 1-form gauge theory with Dirac and complex scalar fields.
It is also gratifying to state that our present theoretical technique has been applied to {\it interacting} non-Abelian
1-form gauge theory with Dirac fields and we have established the absolute anticommutativity of the conserved and off-shell nilpotent
(anti-)BRST charges [43]. 
Presently, we are intensively involved with the above mentioned unsolved problems and we plan to report our results in our
future publication [44].\\

\noindent
{\bf Acknowledgements:} \\

\noindent
Two of us (i.e. NS and AS) would like to gratefully acknowledge the financial supports
from the BHU-Fellowship and CSIR, Government of India, New Delhi, respectively, under which the present investigation
has been carried out. AS is grateful to the Head, Physics Department, Banaras Hindu University, Varanasi 
for the invitation and hospitality during his {\it summer-visit} to BHU where a part
of the present work was completed.\\

\newpage
\vskip 0.5cm
\noindent
$~~~~~~~~~~~${\bf Appendix A: Absolute Anticommutativity and CF-Type Conditions} \\

\noindent
Here we demonstrate that the requirement of the absolute anticommutativity between the BRST and anti-BRST symmetry transformations
leads to the derivation of one of the CF-type restrictions in (7). For instance, it can be checked that 
the operator form of the anticommutator between BRST and anti-BRST transformations, acting on the gauge field $B_{\mu\nu}$, produces 
the following result, namely;
\begin{eqnarray}
\{s_b,\, s_{ab} \}\, B_{\mu\nu} = \partial_\mu \Big[B_\nu - \bar B_\nu \Big] - \partial_\nu \Big[ B_\mu - \bar B_\mu \Big],
\end{eqnarray}
which is a non-zero {\it only} when we take into account  the (anti-)BRST symmetry transformations of the equations (5) and (6). 
The requirement of the absolute anticommutativity $\{s_b,\, s_{ab}\} = 0$ between $s_b$ and $s_{ab}$
demonstrate that one (i.e. $B_\mu - \bar B_\mu = - \partial_\mu \phi_1$) of the CF-type restrictions (cf. (7)) is true. Similarly, 
we observe that the following anticommutator (i.e. $\{ s_d, s_{ad}\}$) between the (anti-)co-BRST symmetry transformations
$s_{(a)d}$, namely;
\begin{eqnarray}
\{s_d,\, s_{ad} \}\, B_{\mu\nu} = \varepsilon_{\mu\nu\eta\kappa}\partial^\eta \Big( {\cal B}^\kappa - {\bar{\cal B}}^\kappa \Big),
\end{eqnarray}
is non-zero {\it only} when we use the (anti-)co-BRST symmetry transformations of (3) and (4). However, the requirement of absolute
anticommutativity (i.e. $\{s_d,\, s_{ad}\} = 0$) between the co-BRST and anti-co-BRST symmetry transformations shows that 
{\it one} of the other (i.e. $ {\cal B}_\mu - {\bar{\cal B}}_\mu = - \partial_\mu \phi_2$) CF-type restrictions
of Eqn. (7) is {\it also} true.

It can be explicitly checked that the rest of the anticommutators are zero when they act on the fields $\rho,\, \lambda,\, C_\mu, \, {\bar C}_\mu,
\, \phi_1,\, \phi_2,\, B_\mu,\, {\cal B}_\mu, {\bar B}_\mu, {\cal B}_\mu,\, {\bar {\cal B}}_\mu$ if we take into account the following nilpotent 
(anti-)BRST and (anti-)co-BRST symmetry transformations:
\begin{eqnarray}
&& s_b\,{\bar B}_\mu  =-\partial_\mu \lambda, \quad  s_{ab}\,B_\mu = \partial_\mu \rho
,\quad s_b {\bar{\cal B}}_\mu = 0, \quad s_{ab} {\cal B}_\mu = 0, \nonumber\\
&& s_d\, {\bar{\cal B}}_\mu   = \partial_\mu \rho, \quad s_{ad}\, {\cal B}_\mu = - \partial_\mu \lambda,\quad
s_d {\bar B}_\mu = 0,\quad s_{ab}\,B_\mu = 0,
\end{eqnarray}
besides the proper (anti-)BRST and (anti-)co-BRST symmetry transformations (3), (4), (5) and (6) 
that are listed in Sec. 2. Thus, we have proven that {\it both} conditions of (7) are {\it true} when 
we demand the absolute anticommutativity between the BRST and anti-BRST as well as the co-BRST and anti-co-BRST symmetry 
transformations separately and independently. It is straightforward to check that the CF-type
restriction: $B_\mu - \bar B_{\mu} = - \partial_\mu\phi_1$ is invariant under the (anti-)BRST
symmetry transformations (5), (6) and (A.2). In exactly similar fashion, it can be
verified that the CF-type restriction ${\cal B}_\mu - \bar {\cal B}_\mu = - \partial_\mu \phi_2$
remains invariant under the (anti-)co-BRST symmetry transformations that are listed in Eqns. (3), (4) and (51).

\vskip 0.5cm
\noindent
$~~~~~~~~~~~${\bf Appendix B: Superfield Expansion and Absolute Anticommutativity} \\

\noindent
As has been claimed in our present endeavor, the observation of the absolute anticommutativity property of the nilpotent
and conserved charges is a {\it novel} observation because we have taken into account {\it only} the (anti-)chiral super expansions of the 
(anti-)chiral superfields. To highlight and corroborate this statement, we show (in our present Appendix) that the property of the absolute 
anticommutativity is very {\it natural} when we take into account the {\it full} expansion of the superfields
(defined on the ($D, 2$)-dimensional supermanifold) along the Grassmannian directions ($\theta,\, \bar\theta$).
Towards this goal in mind, let us start with the {\it full} expansion of the generic superfield ${\tilde \Phi} (x, \theta, \bar\theta)$ 
(defined on the $(D, 2)$-dimensional supermanifold) corresponding to its ordinary counterpart $\phi(x)$ 
(defined on the $D$-dimensional ordinary Minkowskian spacetime manifold) as
\begin{eqnarray}
{\tilde \Phi} (x, \theta, \bar\theta) = \phi(x) + \theta\,  P(x) + \bar\theta\, Q(x) + i\, \theta\bar\theta\, M(x),
\end{eqnarray}  
where the fields $P(x), Q(x)$ and $M(x)$ are the secondary fields. If $\phi(x)$ is fermionic field, then, the pair of fields ($P(x), Q(x)$) would be bosonic
and $M(x)$ would be fermionic. However, if  $\phi(x)$ were bosonic, then, the pair ($P(x), Q(x)$) would be fermionic 
and the field $M(x)$ is bosonic. These conclusions are drawn because of the fermionic nature of the Grassmannian 
variables: $\theta^2 = \bar\theta^2 = 0, \; \theta \bar\theta + \bar\theta \theta = 0$. It is straightforward to note that: 
\begin{eqnarray}
 \frac{\partial}{\partial \theta}\, \frac{\partial}{\partial \bar\theta}\; {\tilde \Phi} (x, \theta, \bar\theta) = - i\, M(x), \qquad \quad
 \frac{\partial}{\partial \bar\theta} \,\frac{\partial}{\partial \theta}\; {\tilde \Phi} (x, \theta, \bar\theta) =  i\, M(x).
\end{eqnarray}  
In other words, we obtain the following (from the above relationships), namely; 
\begin{eqnarray}
\Bigl(\frac{\partial}{\partial \theta}\, \frac{\partial}{\partial \bar\theta} +  \frac{\partial}{\partial \bar\theta}\, \frac{\partial}{\partial\theta}
\Bigr) \; {\tilde \Phi} (x, \theta, \bar\theta) = 0.
\end{eqnarray}
As pointed out earlier, we have taken here the {\it full} expansion of the $(D, 2)$-dimensional 
superfield ${\tilde \Phi} (x, \theta, \bar\theta)$ along the Grassmannian directions ($\theta, \bar\theta$) of the
($D, 2$)-dimensional supermanifold. This is why, we have obtained the interesting relationship (55) which establishes:
$\partial_\theta\, \partial_{\bar\theta} + \partial_{\bar\theta}\, \partial_\theta = 0$.

In the main body of the text, we have proven that the following relationships between the nilpotent ($s^2_r = 0$) 
symmetry transformations ($s_r$) and the translational generators ($\partial_\theta, \, \partial_{\bar\theta}$), on the
$(D, 2)$-dimensional supermanifold, are true
\begin{eqnarray}
&&\frac{\partial}{\partial \theta}\;  {\tilde \Omega}^{(r)} (x, \theta, \bar\theta)\Big|_{\bar\theta = 0} = s_r\,\omega (x)
\equiv \mp \, i\,[\omega(x), Q_r]_\pm,
 \quad \quad
r= ab, ad,\nonumber \\
&&\frac{\partial}{\partial \bar\theta}\;{\tilde \Omega}^{(r)} (x, \theta, \bar\theta)\Big|_{\bar\theta = 0} = s_r\,\omega (x)
\equiv \mp \, i\,[\omega(x), Q_r]_\pm, \quad \quad
r= b, d,
\end{eqnarray}
where ${\tilde \Omega}^{(r)} (x, \theta, \bar\theta)$ corresponds to the generic
superfield (obtained after the application of BRSTIRs and CBRSTIRs) and $\omega(x)$ is the generic ordinary
field (defined on the $4D$ Minkowskian spacetime manifold). Here the symbols $s_r$ and $Q_r$ denote the (anti-)BRST
and (anti-)co-BRST symmetry transformations and corresponding conserved charges. Thus, it is clear that we have the 
following mappings amongst ($s_{(a)b}, \, s_{(a)d}, \, Q_{(a)b},\, Q_{(a)b},\, \partial_\theta, \partial_{\bar\theta}$):
\begin{eqnarray}
&&\lim_{\theta = 0} \,\frac{\partial}{\partial \bar\theta} \quad \Longleftrightarrow \quad (s_b,\, s_d) \qquad
\Longleftrightarrow  \quad (Q_b, Q_d), \nonumber \\
&&\lim_{\bar\theta = 0} \,\frac{\partial}{\partial \theta}\quad \Longleftrightarrow \quad (s_{ab},\, s_{ad}) \quad \;
\Longleftrightarrow \quad (Q_{ab}, Q_{ad}). 
\end{eqnarray}
With the above identifications, it is clear that, in their operator forms, we have the following key
correspondence from (55) and (57):
\begin{eqnarray}
&&\hspace{- 1.4cm} (\partial_\theta \,\partial_{\bar\theta} + \partial_{\bar\theta}\,\partial_\theta = 0) 
\;\Longleftrightarrow\; (s_b \,s_{ab} + s_{ab}\, s_b = 0, \;  s_d \,s_{ad} + s_{ad}\, s_d = 0) \quad \mbox { also}\nonumber \\
&& \hspace{- 1.4cm}(\partial_\theta \,\partial_{\bar\theta} + \partial_{\bar\theta}\,\partial_\theta = 0)\Longleftrightarrow  
(Q_b \,Q_{ab} + Q_{ab}\, Q_b = 0, \;  Q_d \,Q_{ad} + Q_{ad}\, Q_d = 0).
\end{eqnarray}
We draw the conclusion that the absolute anticommutativity property is very natural when we have
the {\it full} expansion of the superfields along ($\theta, \bar\theta $)-directions of 
the $(D, 2)$-dimensional supermanifold. This happens because of the identification
of the Grassmannian translational generators ($\partial_\theta, \, \partial_{\bar\theta}$) with
the nilpotent symmetry transformations (and corresponding conserved and nilpotent charges). We further
note that the identifications in (57) also imply the following
absolute anticommutativity, namely;
\begin{eqnarray}
&&\hspace{- 1.4cm}(\partial_\theta \,\partial_{\bar\theta} + \partial_{\bar\theta}\,\partial_\theta = 0) 
\;\Longleftrightarrow\; (s_b \,s_{ad} + s_{ad}\, s_b = 0, \;  s_d \,s_{ab} + s_{ab}\, s_d = 0) \quad \mbox { also} \nonumber \\
&&\hspace{- 1.4cm}(\partial_\theta \,\partial_{\bar\theta} + \partial_{\bar\theta}\,\partial_\theta = 0) \Longleftrightarrow  
(Q_b \,Q_{ad} + Q_{ad}\, Q_b = 0, \;  Q_d \,Q_{ab} + Q_{ab}\, Q_d = 0).
\end{eqnarray}
The above relations automatically leads to: $\{s_b, s_{ad}\} = 0, \; \{s_d, s_{ab}\} = 0$
as well as $\{Q_b, Q_{ad}\} = 0, \; \{Q_d, Q_{ab}\} = 0$.

\end{document}